\documentclass[useAMS,usenatbib]{mn2e}
\usepackage{graphicx}
\usepackage{amsmath}
\usepackage[usenames]{color}
\usepackage[T1]{fontenc}
\usepackage{aecompl}

\def\L{\emph{ln-}$\mathcal{L}$}

\def\mearth{\hbox{M$_{\oplus}$}}
    \title[Systematic effects in HARPS-N line profile measurements]
   {High-cadence spectroscopy of M-dwarfs.\\
    I. Analysis of systematic effects in HARPS-N line profile measurements on the bright binary GJ~725A+B}
   
     \author[Berdi\~nas et. al.]{
          Z. M. Berdi\~nas,$^{1,2}$\thanks{E-mail: zaira@iaa.es}
     	 P. J. Amado,$^{1}$	
   	 G. Anglada-Escud\'e,$^{3,4}$
	 C. Rodr\'iguez-L\'opez,$^{1}$
	 J. Barnes$^{5}$\\
	 $^{1}$Instituto de Astrof\'isica de Andaluc\'ia -CSIC, Glorieta de la Astronom\'ia S/N, 
	 E-18008 Granada, Spain\\
	 $^{2}$Universidad de Granada-PhD Program in Physics and Mathematics (FisyMat), 18071 Granada, Spain\\
	 $^{3}$School of Physics and Astronomy, Queen Mary University of London, 327 Mile 
	 End Rd., London, E1 4NS, UK\\
	 $^{4}$Centre for Astrophysics Research, University of Hertfordshire, College Lane,
	 Hatfield, Herts AL10 9AB, UK\\
	 $^{5}$Department of Physical Sciences, The Open University, Walton Hall, Milton Keynes, MK7 6AA, UK
	 }

\begin{document}
   
    \date{Accepted YYYY MM DD. Received MM DD, YYYY; in original form MM DD, YYYY}

\pagerange{\pageref{firstpage}--\pageref{lastpage}} \pubyear{2015}
\maketitle
\label{firstpage}

%________________________________________________________________
\begin{abstract}
%________________________________________________________________
Understanding the sources of instrumental systematic noise is a must to improve the design of future spectrographs. In this study, we alternated observations of the well-suited pair of M-stars GJ~725A+B to delve into the sub-night HARPS-N response. Besides the possible presence of a low-mass planet orbiting GJ~725B, our observations reveal changes in the spectral energy distribution (SED) correlated with measurements of the width of the instrumental line profile and, to a lower degree, with the Doppler measurements. To study the origin of these effects, we searched for correlations among several quantities defined and measured on the spectra and on the acquisition images. 

We find that the changes in apparent SED are very likely related to flux losses at the fibre input. Further tests indicate that such flux losses do not seriously affect the shape of the instrumental point spread function of HARPS-N, but identify an inefficient fitting of the continuum as the most likely source of the systematic variability observed in the FWHM. This index, accounting for the HARPS-N cross-correlation profiles width, is often used to decorrelate Doppler time-series. We show that the Doppler measurement obtained by a parametric least-squares fitting of the spectrum accounting for continuum variability is insensitive to changes in the slope of the SED, suggesting that forward modeling techniques to measure moments of the line profile are the optimal way to achieve higher accuracy. Remaining residual variability at $\sim$1~m/s suggests that for M-stars Doppler surveys the current noise floor still has an instrumental origin.

\end{abstract}


%________________________________________________________________
\section{Introduction}
%________________________________________________________________
The radial velocity method to detect extrasolar planets has now reached
long-term precisions of $\sim$ 1 $\rm m\,s^{-1}$, and new instruments are being
built to achieve accuracies in the few tens of $\rm cm\,s^{-1}$ regime
\citep[e.g. ESPRESSO,][]{pepe2010}. A thorough understanding of the current instruments 
and their standard extraction, reduction and radial velocity determination pipelines 
is essential in achieving their ultimate capabilities, and
identifying the key technical aspects that need refinement in the next
generation of instruments. While 1~$\rm m\,s^{-1}$ precision has been enough to
identify the emerging population of super-Earths in compact planetary systems, and to detect potentially habitable planets around M dwarfs, achieving $\rm sub-m\,s^{-1}$ precision at all time-scales is needed to
efficiently characterize Earth and sub-Earth mass objects in hot-to-warm orbits
around G and K dwarfs. There is active discussion on whether the reported $\sim$ 1$\rm
m\,s^{-1}$ long-term precision limit, measured on stabilised spectrographs, is
caused by stellar processes, instrumental effects or both. While certain stars
are, intrinsically, more Doppler variable than this limit, it is unlikely
that the 0.8~-~0.9~$\rm m\,s^{-1}$ (reported lower noise limit in HARPS-ESO
observations by \citealp{pepe2011}, and of a handful of very quiet M~dwarfs, e.g.
GJ~699 and GJ~588, by \citealp{anglada2012a}) is an universal limit for
Doppler measurements. 

Studying the limits of precision of Doppler spectroscopy in the short period domain of M-type dwarfs is one of the primary goals of the ``Cool Tiny Beats survey" \citep[CTB;][]{anglada2014, berdinas2015}. The survey, started in 2013, was designed to obtain high-cadence observations (continuous observing of a single target during several consecutive nights) of a small sample of M-dwarfs with HARPS-ESO (High Accuracy Radial Velocity Planet Searcher) and HARPS-N \citep{mayor2003, pepe2004, cosentino2012}. Our objective was threefold: (1) the detection of small planets in sub-day period orbits, (2) the search for pulsations, and (3) the study of the time-scales of stellar activity. 

From the analysis of those data, it was obvious that some kind of instrumental or reduction-process effect was producing intra-night and night-to-night variability common to several targets. After the first runs in each instrument, two of such effects were quickly identified: (1) Doppler shift correlations with the signal-to-noise ratio \citep[SNR,][]{bouchy2009}, and (2) $\sim$0.5~-~1.5~$\rm m\,s^{-1}$ night-to-night jumps in the radial velocity (RV) series. Berdi\~nas et al. in prep. (hereafter, CTB 2016) find that the first one was partly due to the Charge Transfer Efficiency effect, and provide an empirical correction that should at least be valid for M-dwarfs observed from HARPS-ESO. The effect was corrected in HARPS-N (but not in HARPS-ESO) by introducing, in 2013, modifications to the standard Data Reduction Software (DRS) pipeline (Lovis, C., Pepe, F., priv. comm.). The second effect was found to be caused by random and systematic errors in the wavelength solution of unknown origin (Lo Curto, G., priv. comm.). CTB 2016 also provides a solution for this effect that consists in using a mean wavelength solution for each observing run instead of the one derived from individual night calibrations, for both HARPS-ESO and HARPS-N.

While these corrections substantially reduced the common systematic variability of several observed stars, yet another source of systematic noise was strongly affecting HARPS-N data of M-dwarf stars. This variability correlates with the airmass, and was found to produce structured Doppler noise at the level of $\sim$2~$\rm m\,s^{-1}$, with even larger amplitude in the full width at half maximum (FWHM) of the cross-correlation function. Similar trends have been recently reported for circular fiber-fed spectrographs like SOPHIE for other spectral types \citep{bohm2015}.

There are many environmental effects which worsen the higher the airmass is, such as the atmospheric refraction, which increases with the air density at low altitudes splitting the star image in its blue-red components in the zenithal direction; the seeing, which also increases at low altitudes, being the blurring more severe for shorter wavelengths; or the atmospheric extinction, caused by the Rayleigh scattering and by the molecules and dust absorption. In terms of airmasses, the atmospheric dispersion dominates below $\sim50\deg$ on nights with good seeing ($\sim$1 arcsecond in the zenith). To correct for this, HARPS-N includes an atmospheric dispersion corrector (ADC). 

The aim of this work is to study the response of the instrument to any of these sources in the intra-night regime, check if these effects lead to variability of the spectra and provide means for correcting them. The paper is organised as follows. We describe the observations and the properties of the targets used for this study (GJ~725A and B) in Section~\ref{sec:observations}. The data reduction and the spectroscopic and image indices to be used in this study are presented in Section~\ref{sec:measurements}. Section~\ref{sec:analysis} comprises the analysis, where, in Section~\ref{sec:corr} we show the common variability detected in the alternating observations between GJ~725A and B, and also we intend to assess statistical correlations between spectroscopic observables and various instrument related quantities. In Section~\ref{sec:experiment} we introduce the least square deconvolution profiles (LSD) and perform two experiments to test the algorithms used in this paper (HARPS-TERRA and DRS data reduction softwares). We propose tentative functions to detrend the line width and the RVs measurements in Section~\ref{sec:detrending}. In Section~\ref{sec:spaprof} we present the analysis of the mean line profiles calculated in the cross-dispersed direction. We discuss a promising $\sim$~2.7 days Doppler signal in GJ~725Bb in Section~\ref{sec:dopplerb}. Finally, Section~\ref{sec:conclusions} present the discussion and  conclusions of this work. 

%________________________________________________________________
\section{Observations}\label{sec:observations}
%__________________________________________________________________

We used the HARPS-N spectrograph at the 3.6\,m Telescope Nazionale Galileo
(TNG), settled at the Roque de los Muchachos Observatory in La Palma. HARPS-N is
nearly a twin of HARPS-ESO. To guarantee that the optics keep their alignment, both are vacuum-sealed thermally stabilized fibre-fed \'echelle spectrographs. HARPS-N input fibre samples 1~arcsecond circular
apertures on the sky, producing high-resolution optical spectra from 383 up to
693~nm. The high resolution R$\sim$120000 (at the central wavelength, see table~6 in \citealp{cosentino2012}), makes the radial velocities (RVs) measured from the spectra the most long-term accurate currently available.

	\begin{table}
	\begin{center}
	\caption{Observational parameters of the 2014 run.}     
	\label{tab:runparam}
	\begin{tabular}{lcccc}\hline
	Nights 				& Night 1 & Night 2 & Night 3 & Night 4 \\ \hline
	$\rm N_{A}$ 			& 27 & 40 & 16 & 18\\
	$\rm N_{B}$ 			& 27 & 21 & 31 & 37\\
	$\rm AM_{max}$ 	        & 1.55 & 1.64 & 1.59 & 1.65 \\
	$\rm Seeing$               	& 0.86 & 0.87 & 0.74 & 0.78 \\
	$\rm Seeing_{min}$ 		& 0.58 & 0.60 & 0.48 & 0.59 \\
	$\rm Seeing_{max}$ 		& 2.83 & 1.47 & 3.64 & 1.20 \\
	$\rm SNR_{A}$ 	        & 78 & 54 & 59 & 71  \\
	$\rm SNR_{B}$ 	        & 63 & 49 & 53 & 59  \\\hline
	\end{tabular}   
	\end{center}
	Main observational parameters of the 2014 run (2nd to 5th July). Subindices A and B refer to targets GJ~725A and GJ~725B. $\rm N$ indicates the number of exposures. The SNR and seeing values are median measurements of the nights. SNR values correspond to the spectral order 60, centered at 631 nm. Since the TNG seeing monitor was out of order, we show the values given by the nearby Isaac Newton Telescope. 1.17 was the minimum airmass (AM) of all nights.
	\end{table}
	
Our targets were the bright nearby stars: GJ~725A and GJ~725B\footnote{Check
http://www.pas.rochester.edu/$\sim$emamajek/spt/M3V.txt and
http://www.pas.rochester.edu/$\sim$emamajek/spt/M3.5V.txt for a thorough review
of both stars properties.}. We selected this binary system for being a bright,
\citep[$V$=8.91, $V$=9.69, for A and B respectively,][]{jenkins2009, cutri2003},
nearby \citep[d=3.57 pc, d=3.45 pc,][]{anderson2012}, and common proper motion
pair of almost identical spectral type \citep[M3V and M3.5V,][]{reid1995} with
no previous reports of intense magnetic activity or planets. At epoch 2000,
their projected separation was 13.3~arcseconds, implying a minimum separation of
47.5~AU \citep[Hipparcos,][]{vanleeuwen2007}. 

We observed the stars in two high-cadence different runs, one in 2013 and
another one in 2014. GJ~725A was the primary target of our five-day long,
August 2013 run, whereas, in 2014, GJ~725A and B were observed every night for 4
consecutive nights alternating exposures between the two of them. This gave us a total
of 314 A data points in 2013 and 101 for A and 116 for B in 2014, after
rejecting those with SNR below 45 and airmass above 2.5 (the HARPS-N atmospheric corrector does not work for higher airmasses, F. Pepe, private comm.). In the 2014 run, except for
the first night, we halted the alternating sequence between A and B at the
meridian passage, observing instead only one of the stars. We did this
to improve the cadence of each target at that specific moment of the night.
Exposure times were set to be between 240 and 480~s. The main observational
parameters of the run are shown in Table~\ref{tab:runparam}.

%________________________________________________________________
\section{Measurements}\label{sec:measurements}
%________________________________________________________________
For each observation, we obtained a number of measurements (or indices) to perform this study. These indices are detailed below.

%________________________________________________________________
\subsection{Spectroscopic measurements}

Spectra were processed and extracted with the standard HARPS-N/TNG Data
Reduction Software (DRS). We obtained the RV
measurements using the maximum likelihood matching technique
implemented in the HARPS-TERRA software \citep{anglada2012a}.

The DRS cross-correlates the observed spectra
with a weighted binary mask formed by ones in the theoretical stellar lines positions and widths and by zeros elsewhere \citep{pepe2002}. The resulting cross-correlation function (CCF) acts
as a proxy for the mean line profile for each spectral order. In our study, we will use 
the so-called FWHM index (full width at half maximum of the
Gaussian fitted to the cross-correlation profile). Since the line width might be sensitive
to the presence of spots or magnetic activity \citep[e.g.][]{reiners2013}, this
FWHM is often used as an activity index to decorrelate Doppler
variability. 

%________________________________________________________________
\subsection{Spectral energy distribution measurements (pSEDs and $K$ index)} \label{sec:sed}

\begin{figure}
	\centering
	\includegraphics[width=\columnwidth]{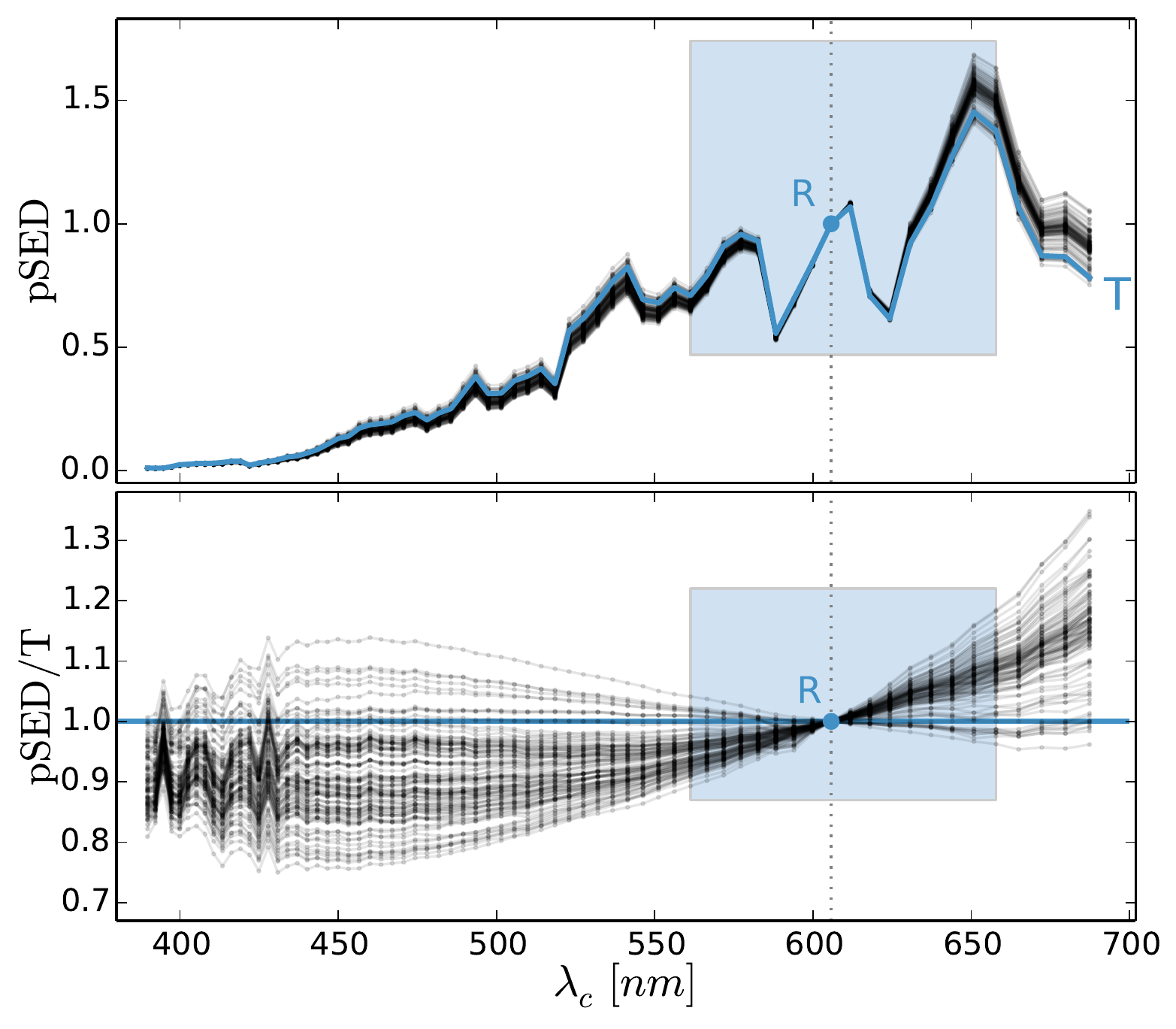}
	\caption{(Upper panel) Sum of the flux of the 69 spectral orders of GJ~725A (the so-called pseudo spectral energy distribution, pSED) plotted versus wavelength ($\lambda_{c}$ refers to the central wavelength of each spectral order). The pSEDs cross the normalization point (R). The blue line T indicates the initial pSED of the run. (Lower panel) Relative flux compared to T. We limited the study to a linear pSED/T region of $\pm8$ spectral orders  from R (blue area). Fluxes were corrected from atmospheric extinction.}
	\label{fig:sed}
	\end{figure}

We also studied the distribution of the flux across the spectra. 
With this aim, we used the spectra as given by the DRS reduction pipeline for each spectral order (the so-called e2ds data-products), and we calculated how the total flux on each spectral order varied relative to a reference order. 
To do this, we firstly added the flux of all pixels in each spectral order, and secondly, we normalised these measurements to the total flux at the spectral order 56, which was selected for having low telluric contamination. We named this 
``normalized pseudo spectral energy distribution" (pSED). As we 
see for the 2014 data of GJ~725A in the upper panel of Figure~\ref{fig:sed}, 
the pSED pivots around the normalization order (R). Changes relative to the 
first co-added spectrum (T) are shown in the lower panel of Figure~\ref{fig:sed}. 

The pSED is expected to vary for a number of environmental effects coupled with the
instrument, as for example the atmospheric dispersion or the chromatic seeing. Account for the atmospheric dispersion is the ADC role, however the flux reaching the telescope will also change as a function of airmass due to the atmospheric extinction (i.e. due to the Rayleigh scattering and to the absorption from molecules and dust in the Earth atmosphere). We corrected the pSED due to atmospheric extinction using the method outlined by
\cite{hayes1975}\footnote{\label{footnote:atmext}See ORM atmospheric extinction
values for a dust-free atmosphere at
http://www.ing.iac.es/Astronomy/observing/manuals/ ps/tech\_notes/tn031.pdf. To
correct from aerosol scattering we used the V-band extinction measurements for
the month of July from table~1 of \cite{garcia-gil2010}.}. Although the
atmospheric extinction is wavelength and airmass dependent, the pSED varied less
than a 4.5 per cent and the wavelength dependence in the pSED remained equally
strong after the correction. 

To quantify the impact of changes in the pSED, we defined a spectroscopic chromatic index
($K$). This index measures the relative changes in the slope of the pSED as a function of wavelength considering that a linear function fits the pSEDs in the surroundings of R (orders inside the blue box of Figure~\ref{fig:sed})
$K$ is obtained from:
		\begin{equation}
		\mathrm{pSED}_{\rm t}\,(\lambda) = [1 + K_{\rm t}\,(\lambda_{\rm{c}} - \lambda_{\rm{cR}} )]\, \rm T(\lambda), 
		\label{eq:k}
		\end{equation}
where $t$ indicates the observations, $\lambda_{\rm{cR}}$ is the
central wavelength of the reference order 56, $\lambda_{\rm{c}}$
refers to the central wavelength of any other order,
$\rm{T}\,(\lambda)$ corresponds to the initial
$\rm{pSED}_{\rm{t_{0}}}(\lambda)$ for each star; and finally,
${K}_{\rm{t}}$ is a coefficient with units of 1/$\lambda$,
that we call chromatic index.

%________________________________________________________________
\subsection{Measurements on the autoguide camera images (Semimajor axis module, $\Phi$, and Circularity index, $\Theta$)}\label{sec:images}

The images obtained by the autoguide camera were additional products that were
used for trying to find an explanation for the systematics. In HARPS-N, the star
light that is not injected into the fibre is reflected towards the autoguide
camera. To do this, the fiber head is fitted in the central hole of a slightly tilted flat mirror. Therefore, the images recorded by this camera show the shape of the star
at the fibre entrance. Two kind of images are stored, the acquisition images,
taken with the star off-fibre before the science exposure, and the autoguide
images, which are the average of the short images ($\sim 100~\rm ms$) taken during the
science exposure to guide the telescope. 

	\begin{figure}
	\centering
	\includegraphics[width=\columnwidth]{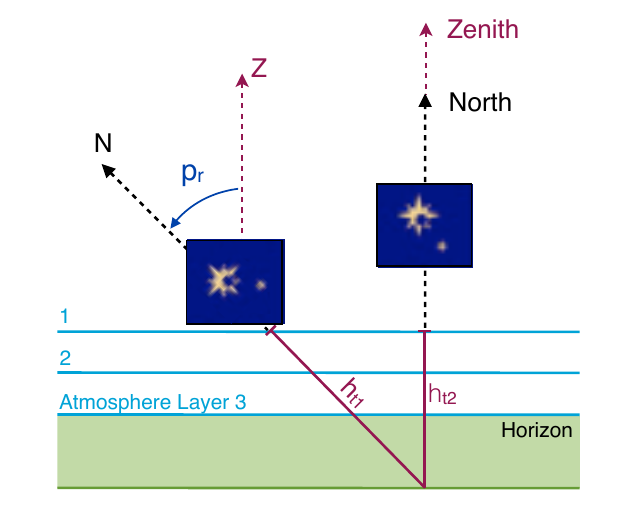}
	\caption{Schematic view of the sky orientation of a non-de-rotated field of view in an altazimuthal telescope like HARPS-N. During the night, the field rotates following the celestial north axis direction (N), while the zenithal axis (Z) remains perpendicular to the atmosphere and does not rotates with the field. The atmosphere chromatically disperses the light more efficiently when the atmospheric layers are thicker (e.g. for $\rm h_{t_{1}}$ compared with $\rm h_{t_{2}}$). $p_{r}$ refers to the parallactic angle, the angle subtended by Z and N.}
	\label{fig:adc}
	\end{figure}

It must be realised that, in an equatorial telescope (or in a de-rotated field of view 
in an altazimuthal telescope), the direction of the dispersion of the atmosphere
changes during the night (following the zenithal direction), whereas in a
non-de-rotated field of view (the case of HARPS-N@TNG) the zenithal
direction remains constant while the field rotates (see Figure~\ref{fig:adc}).

	\begin{figure}
	\centering
	\includegraphics[width=\columnwidth]{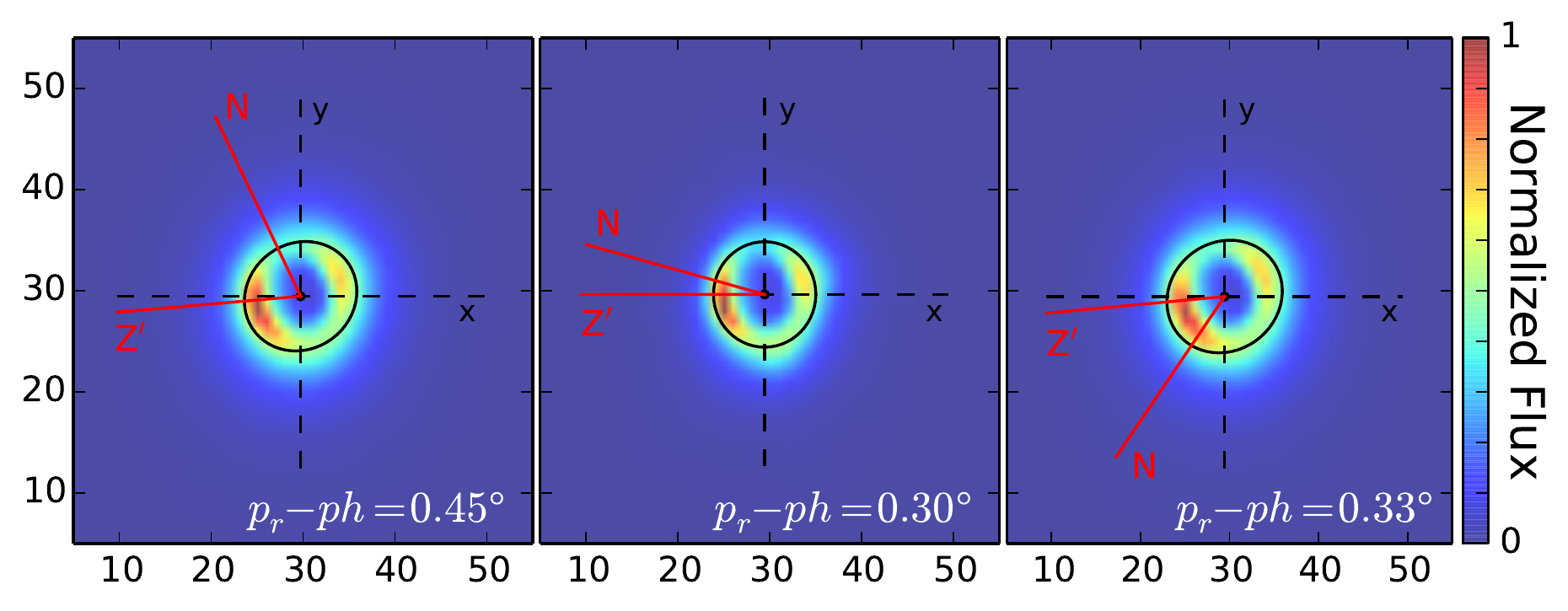}
	\caption{GJ~725A autoguide camera images taken at the beginning (left), close to the meridian crossing (middle), and at the end (right) of the 2nd July 2014 night. The central hole, which corresponds to the fibre position, seems to be elliptical due the slight tilt of the flat mirror which diverts the light towards the autoguide camera. The solid black lines refers to the elliptical fits performed to account for the elongation of the images. After accounting for the field rotation and the fix angle of the derotator, we measured a good agreement of $p_{h}$ --the angle subtended by the north axis N, and the ellipse semimajor axis Z'-- with $p_{r}$, the real parallactic angle (or angle subtended by the north-zenith axis). The distortion of the images is aligned with the atmospheric dispersion axis (Z), an thus, with the airmass.}
	\label{fig:autoguide}
	\end{figure}
	
It is important to note that, as a result of the flat mirror tilt, a stable distortion of the images was expected; i.e. we expected the images to be elongated in the same proportion and direction as we measured for the circular hole of the mirror (see the shape of the central hole in Figure~\ref{fig:autoguide}). However, the autoguide images presented a variable elliptical elongated shape of the star that did not fit the distortion of the mirror hole and did not rotate with the field (within a maximum error of $4^{\circ}$), keeping always aligned with the
zenithal direction. This result, together with the correlation of the elongation with
the airmass (see three autoguide images taken at the beginning, middle and end of
one night, with airmasses 1.30, 1.17, and 1.33 respectively, in
Figure~\ref{fig:autoguide}), led us to hypothesize that the elongated shape was
produced due to an insufficient correction of the atmospheric chromatic
dispersion by the ADC. To validate our hypothesis, we fitted an ellipse to the autoguide images (only the outermost pixels with fluxes within $\pm$1 per cent of the half value between the maximum and the background were considered), and we checked that the elongation was compatible with the zenithal axis. In particular, we measured maximum differences of only $7.8^{\circ}\pm6.2^{\circ}$ between $p_{r}$, the real parallactic angle, and $p_{h}$, its analogous in case the semimajor axis of the ellipse fitted to the image and the zenithal direction were coincident. Furthermore, if this is so, the resulting image is consistent with a superposition of wavelength-dispersed images of the star caused by the atmosphere.
%________________________________________________________________
\subsection{Summary of used measurements}
For the sake of clarity, we summarise here the measurements that will be use in this study.
	\begin{itemize}
	\item{Spectra}
		\subitem RV: Extracted with HARPS-TERRA.  
		\subitem FWHM: Given by the DRS pipeline.  
	\item{Spectral Energy Distribution}
		\subitem \emph{Spectroscopic Chromatic Index} ($K$): Quantifies the changes in the slope of the SED. 
	\item{Autoguide images}
		\subitem \emph{Semimajor axis module} ($\Phi$): Module of the semimajor axis resulting from fitting an ellipse to the autoguide images.
		\subitem \emph{Circularity index} ($\Theta$): Seeing difference, in arcseconds, between the x- and y-axis, obtained from the autoguide image headers. 

	\end{itemize}

%________________________________________________________________
\section{Analysis} \label{sec:analysis}%________________________________________________________________

\subsection{Systematic effects in the time-series} \label{sec:corr}
%________________________________________________________________

\begin{figure}
	\centering
	\includegraphics[width=\columnwidth]{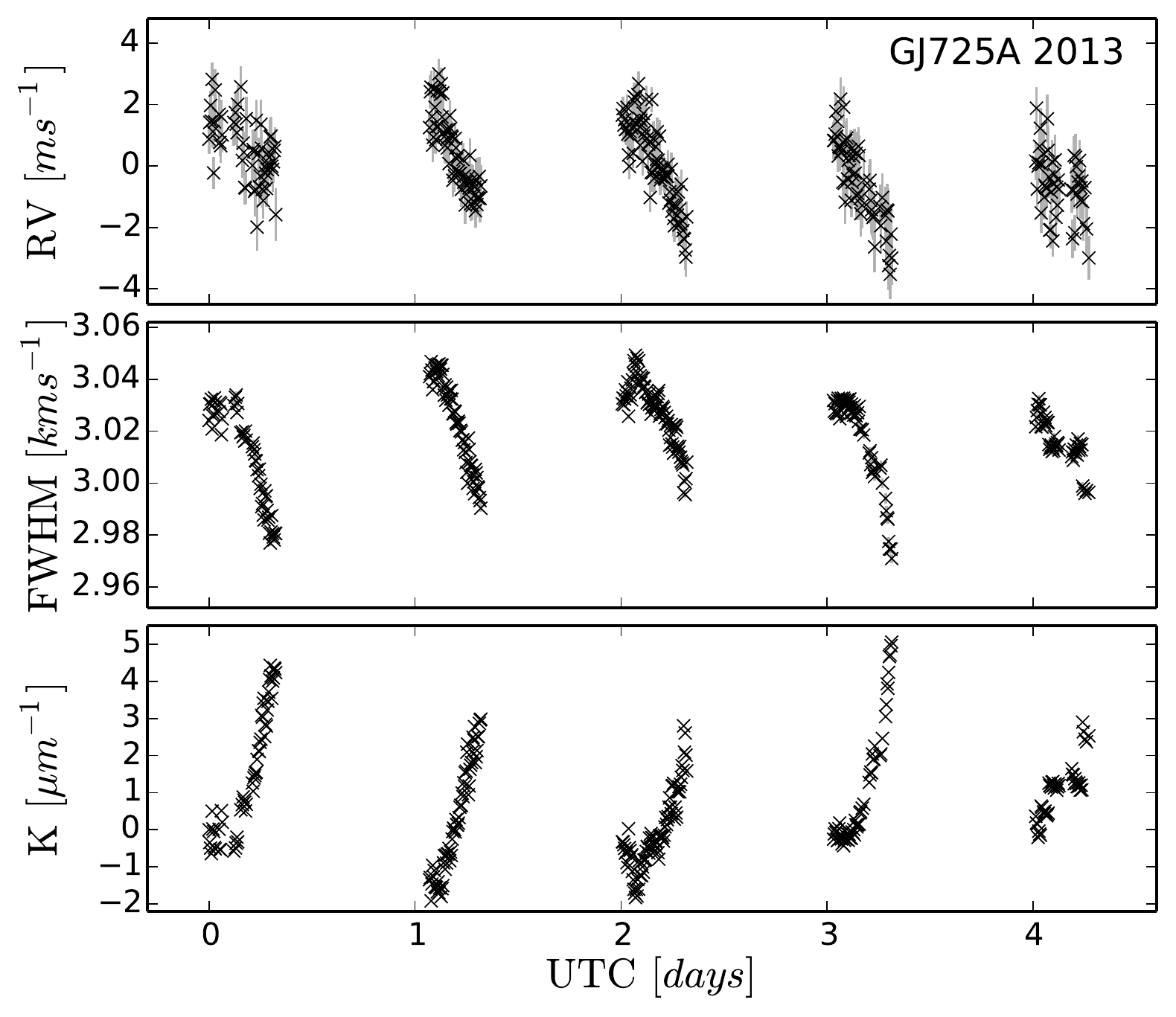}
	\caption{RV (top), FWHM (middle), and $K$ (bottom) time-series of
GJ~725A observed during five nights run of 2013. The FWHM vary tens of $\rm
m\,s^{-1}$ peak-to-peak, while the RV peak-to-peak variability is $\sim$
4.5~m\,s$^{-1}$.}
	\label{fig:run13timeser}
	\end{figure}

GJ~725A was the primary target of the high-cadence 2013 run. These observations showed a
clear $\sim$~1 day Doppler signal highly correlated with the airmass and common to the RVs and the FWHM time-series (see Figure~\ref{fig:run13timeser}). We suspected an instrumental origin, but with only one observed star, we could not distinguish systematic effects from astrophysical ones.

As a consequence, we designed a second observational run in 2014 to separate
astrophysical from instrumental effects: given that GJ~725A+B is a
common proper motion pair of almost identical components, obtaining alternating
observations of both components is a simple and efficient way to point out common trends. 
For short periods of time, we also halted the alternation strategy to monitor a single star and
increase sensitivity in the high frequency domain.

\begin{figure}
	\centering
	\includegraphics[width=\columnwidth]{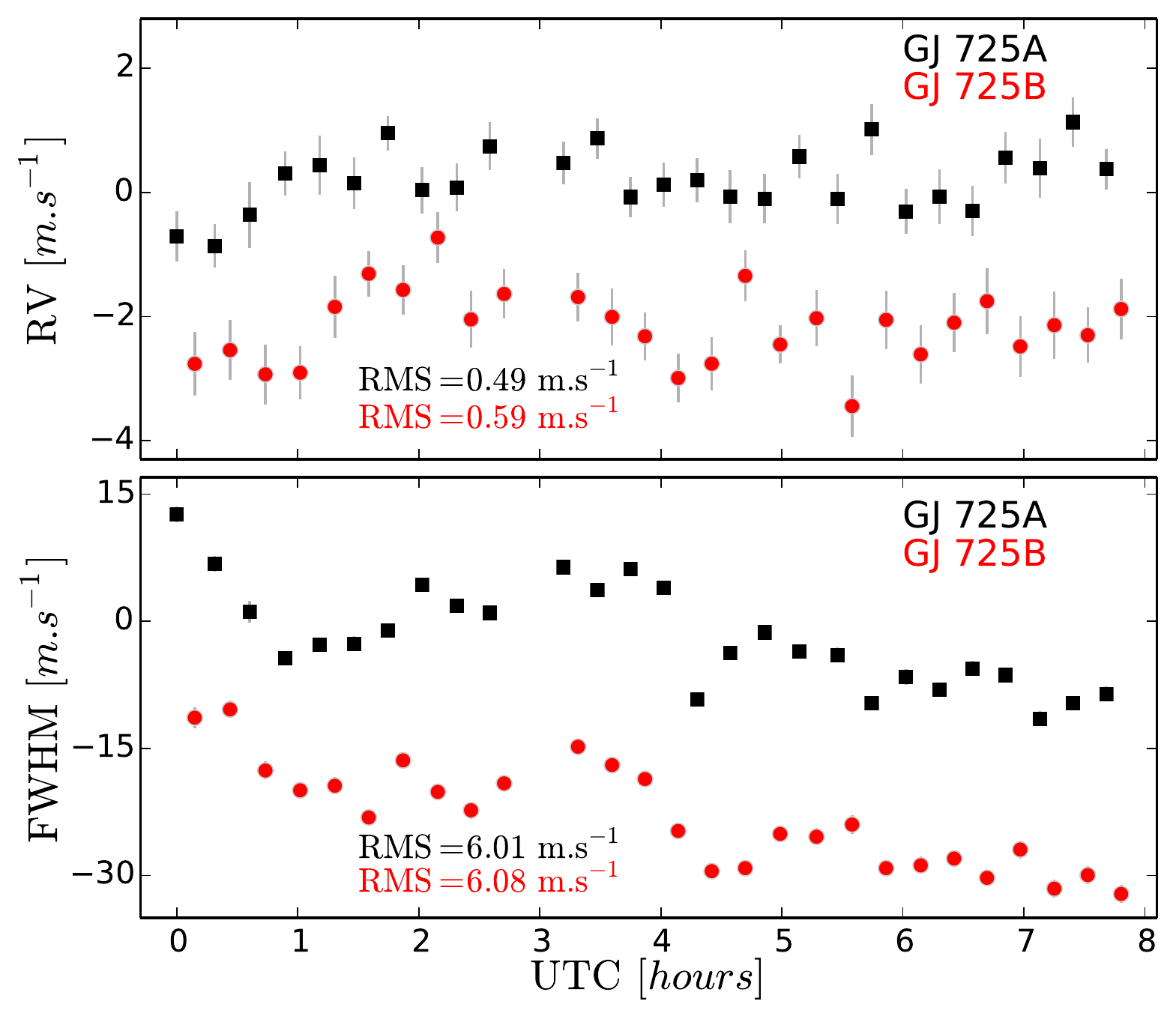}
	\caption{GJ~725A (black squares) and GJ~725B (red
dots) RVs (upper panel) and FWHM (lower panel) time-series for
the first HARPS-N run night on 2nd July 2014. For the sake of
clarity, we subtracted the FWHM mean velocities (2989.97 and
2975.62~$\rm m\,s^{-1}$ for GJ~725A and B, respectively), and
we shifted the FWHM (-20~$\rm m\,s^{-1}$) and the RVs
(-2.3~$\rm m\,s^{-1}$) of GJ~725B. We referred the time-axis
to the beginning of the night at 21h 11m 28s UTC. Both stars 
show common variability in both quantities confirming
an instrumental origin.}
	\label{fig:dataproducts}
	\end{figure}

Correlated variability, common to both stars, can readily be spotted by direct
inspection of the time-series (see Figure~\ref{fig:dataproducts}). For the first
observed night, the RV correlation coefficient between GJ~725A and B time series
was $r=0.30$ and $r=0.87$ for the FWHM. 

	\begin{figure}
	\centering
	\includegraphics[width=\columnwidth]{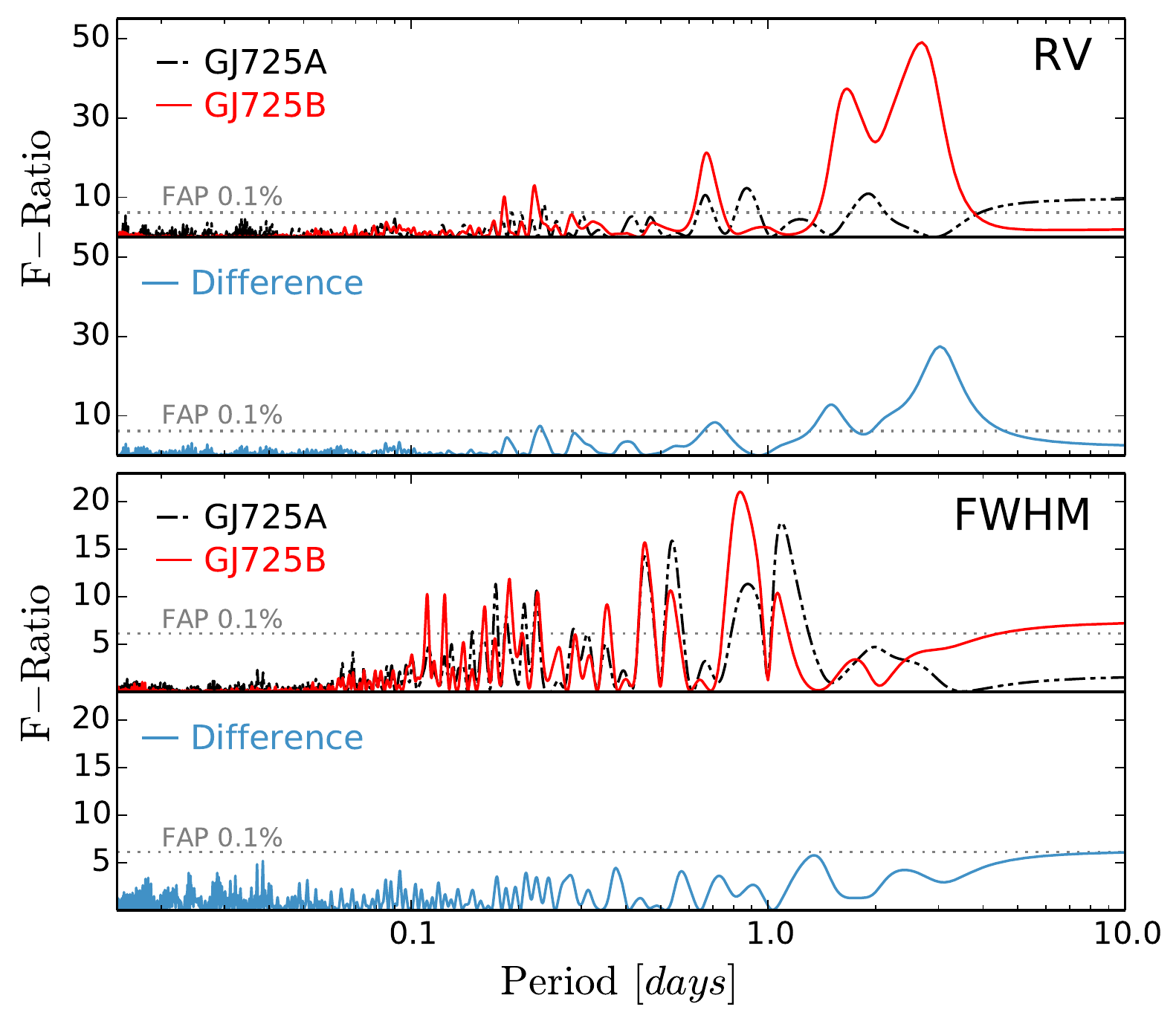}
	\caption{Periodograms for the RV (upper panel) and the FWHM (lower
panel) for the GJ~725A (black line) and GJ~725B (red line) data of the 2014 run.
The blue line corresponds to the periodogram of the two stars differential
time-series. Note the $\sim3$~d peak in the RVs difference over the 0.1 per cent
false alarm probability threshold. We avoided sampling effects by interpolating the GJ~725B  times to
the GJ~725A ones. We also excluded high-cadence observations. Common
variabilities result in common peaks, which disappear on the differential
periodograms. The remaining peaks are real GJ~725A or GJ~725B signals.}
	\label{fig:difrunper}
	\end{figure}

\begin{figure}
	\centering
	\includegraphics[width=\columnwidth]{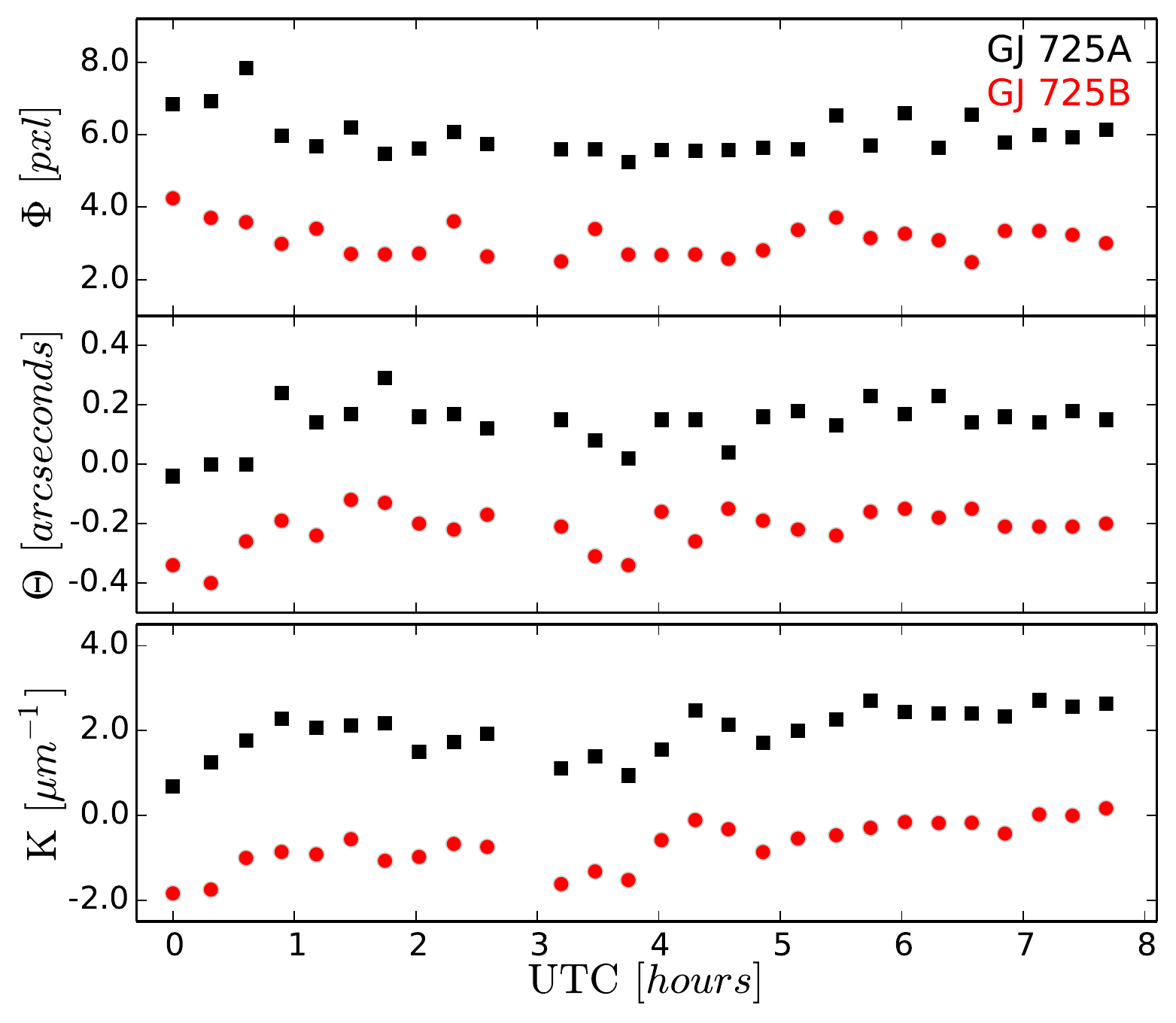}
	\caption{Time-series of $\Phi$, $\Theta$ (images indices), and $K$
(spectroscopic chromatic index) for the 2nd of July 2014. Black squares and red
dots correspond to GJ~725A and GJ~725B, respectively. For the sake of clarity,
we shifted GJ~725B -3~pxl for $\Phi$, -0.35~arcseconds for $\Theta $, and -2.5
$\mu m^{-1}$ for $K$. The time axis is referred to the first observation.}
	\label{fig:indices}
	\end{figure}

To isolate real signals arising from either star, we computed periodograms of
the time-series of the individual stars, together with periodograms of their
difference (see Figure~\ref{fig:difrunper}). 
Periodograms in Figure~\ref{fig:difrunper} follow the procedures outlined in \cite{zechmeister2009}. That is, we used the F-ratio statistic to find what is the period of the sinusoid best-fitting the data \citep[see detailed description in][]{zechmeister2009}.
We used these periodograms because, while being formally equivalent to the Lomb--Scargle periodograms \citep{scargle1982}, they are less susceptible to aliasing and provide more accurate frequencies.
To compute the differential periodograms, we calculated the difference between the GJ~725A and B velocities, evaluating B at the A observing
epochs (we linearly interpolate the GJ~725B time-series). RV and FWHM periodograms of
the individual stars showed common peaks (see black and red lines in Figure~\ref{fig:difrunper}), 
and most of these peaks disappear in the differential periodograms (blue lines). This result
suggests common instrumental variability on the data-products of both stars. Note a
promising signal arising over the 0.1 per cent of the false alarm probability \citep[or FAP,][]{cumming2004} threshold at $\sim$3~d in the RV differential periodogram. This signal may corresponds to some real variability on GJ~725B as we discuss later in Section~\ref{sec:dopplerb}. The indices calculated over the autoguide images and the pSED (e.i. $\Phi$, $\Theta$, and $K$) also show common variability on both stars as we show in Figure~\ref{fig:indices}.

%________________________________________________________________
\subsubsection{Analysis of the correlated variability between indices}

We analysed all possible correlations between our spectroscopic, autoguide and
pSED indices using the following procedure. Given two indices $x_i$ and $y_i $,
we modeled the existence of correlations by fitting a linear relation between them:
$y_{i}=a+bx_{i}$, where $a$  and $b$ are the free parameters. Assuming that the
null model containing no correlation is $b=0$ and $a = \bar{y}$, 
our significance assessment consists in obtaining the improvement in the
$\chi^2$ statistic as:  

\begin{equation} 
  \Delta\chi^{2}=   \sum_{i}\left (
  \frac{y_{i}-\left(a+bx_{i}\right)}{\sigma_y{_{i}}}\right)^{2} - \sum_{i}\left
  (\frac{y_{i}-\bar{y_{i}}}{\sigma_{y_{i}}}\right)^{2}, 
\end{equation} 

\noindent and then determining whether such an improvement could be caused by a
fortunate arrangement in the noise. Instead of using analytic expressions for
the expected distribution of $\Delta\chi^{2}$ in the presence of noise, we
obtained its empirical distribution using a Monte Carlo approach. For each
correlation under investigation, we generated a large number of synthetic
datasets by randomly swapping the $y_i$ values while keeping the same $x_i$. The
false alarm probability (or FAP) is empirically define here as the number of these random experiments which give a
$\Delta\chi^{2}$ larger than the one obtained from the real data divided by the
total number of experiments. We performed enough experiments (in all cases $10^{3}-10^{4}$ depending on
the obtained FAP) to ensure that the
FAP estimates were accurate enough.

\begin{figure*}
	\includegraphics[width=\textwidth]{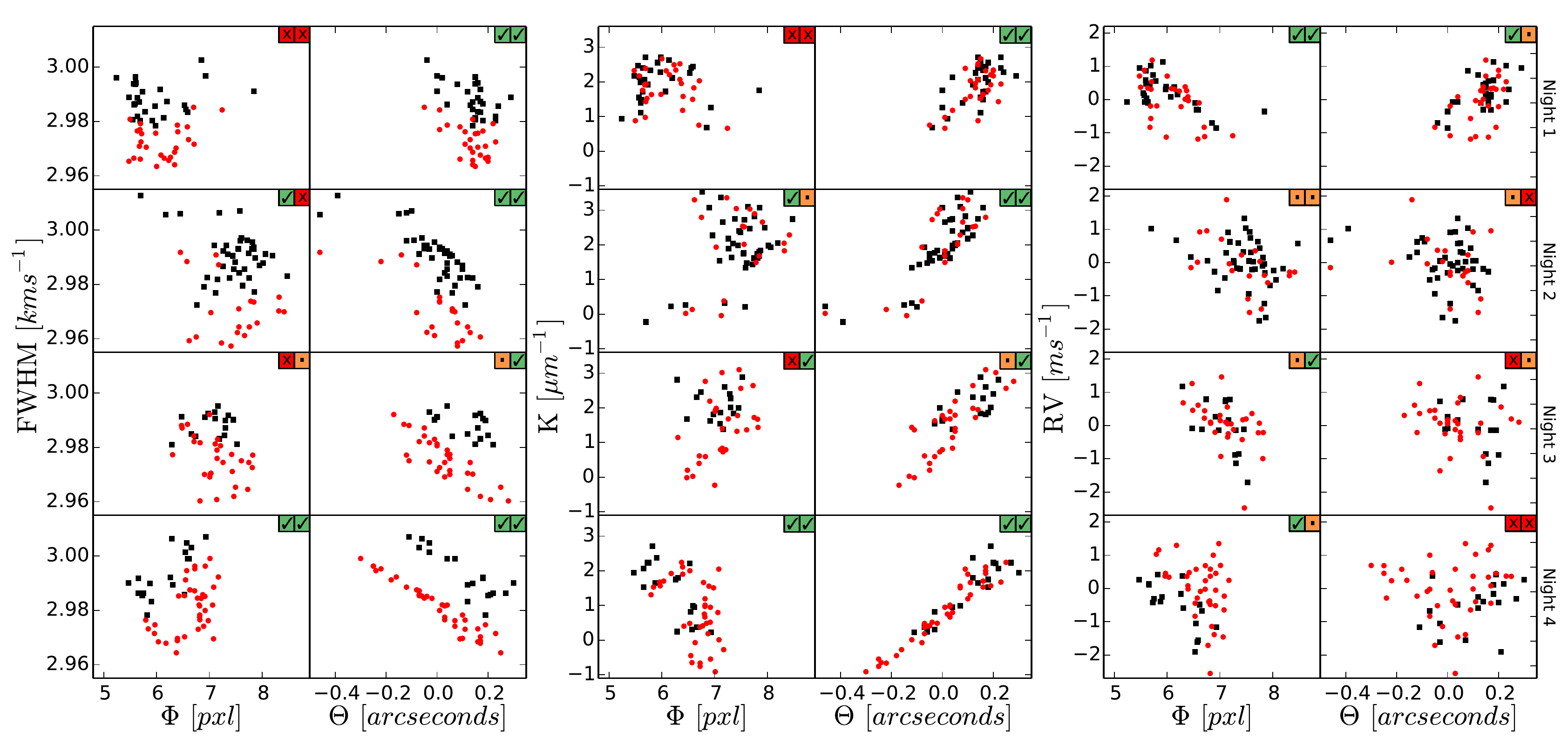}
	\caption{Correlation of the spectroscopic indices, $\rm{FWHM}$ (left panel), $K$ (central panel), and $\rm{RV}$ (right panel), with the image indices, $\Phi$ and $\Theta$. Each row matches one night. Black squares correspond to GJ~725A and red dots to GJ~725B. Two squares in the top right corners of each plot indicate the FAP for GJ~725A and B, respectively from the left: ticked green squares means $\mathrm{FAP} < 1\%$, dotted orange $1\% \leq \mathrm{FAP} < 10\%$ and crossed red $\mathrm{FAP} \geq 10\%$. Main correlations are: FWHM-$\Theta$, $K$-$\Theta$ and RV-$\Phi$. GJ~725B RVs are residuals to the Doppler signal discussed in Section \ref{sec:dopplerb}.}
	\label{fig:corrimag}
	\end{figure*}

\begin{figure*}
	\includegraphics[width=11cm]{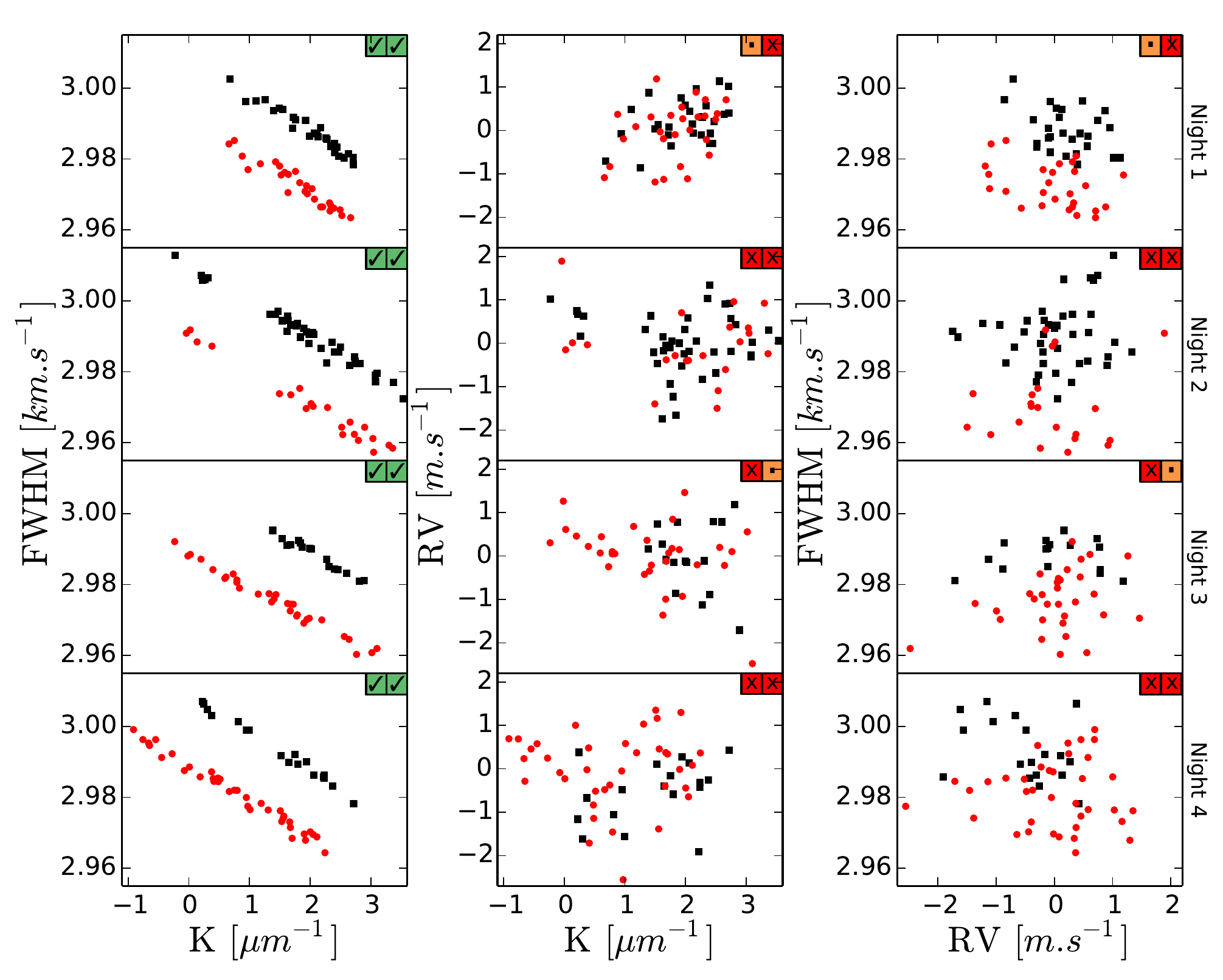}
	\caption{Best correlation linear models fitted for the spectroscopic indices: $\mathrm{FWHM}-K$ (left panel), $\mathrm{RV}-K$ (central panel), and $\mathrm{FWHM}-\mathrm{RV}$ (right panel). Markers and top right corner squares are coded as in Figure~\ref{fig:corrimag}. Among these indices FWHM-$K$ was the only correlation detected.}
	\label{fig:corrspec}
	\end{figure*}

The correlation analysis for the 2014 measurements (which include both
stars) are shown in Figures~\ref{fig:corrimag} and~\ref{fig:corrspec}. The two
squares in the top right corners of the individual plots indicate the FAP
(ticked green squares mean $\rm FAP<~1\%$, dotted orange mean
$1\%\leq\mathrm{FAP}<10\%$, and crossed red mean $\mathrm{FAP}\geq10\%$) for
GJ~725A (black squares data) and GJ~725B (red dots data). We excluded five
outliers on $\Phi$ (three for having a dispersion above 4.5 the measured standard
deviation of the sample, and two produced by sudden seeing increases and
corresponding SNR in the spectra below 40).

The strongest correlations correspond to the FWHM and the spectroscopic $K$-index
(Figure \ref{fig:corrspec}, first column). The correlation is so strong that all
nights tested resulted in compatible fitted parameters for both stars. This
dependence, together with the correlation of $K$ with the image index $\Theta$
(Figure \ref{fig:corrimag}, first panel), seems to indicate that the issues at
the fiber coupling are propagating all the way down to the science spectrum and
affecting its SED and its mean line profile. In other words, the instrumental
profile of the spectrograph seems to vary with wavelength and time. It is important
to note that variable seeing necessarily increases the autoguide image radius
momentarily, but it will increase proportionally in both the x- and y-axis.
As a result, $\Theta$ will remain as a good measurement of the asymmetry in the
fibre injection.

We also measured a less strong correlation between the RVs and the image index
$\Phi$ (Figure \ref{fig:corrimag}, third panel). We used the GJ~725B RVs after subtracting the Doppler signal discussed in Section~\ref{sec:dopplerb}. The correlation is not as strong as that of the FWHM with
$K$, however both stars have the same qualitative behaviour. The $\rm
FAP$ of all observed nights remains below the 10 per cent threshold. 

%________________________________________________________________
\subsubsection{Analysis of the 2013 single star run on GJ~725A} \label{sec:run2013}

Unfortunately, most of the acquisition and some of the guiding images from 2013
were lost due to an error in the acquisition software which was later solved in
March 2014\footnote{\label{fn:technical14}Instrument upgrades can be consulted
in http://www.tng.iac.es/instruments/harps/.}. As a result, we could not obtain
consistent measurements over the autoguide images.

\begin{figure*}
	\includegraphics[width=15cm]{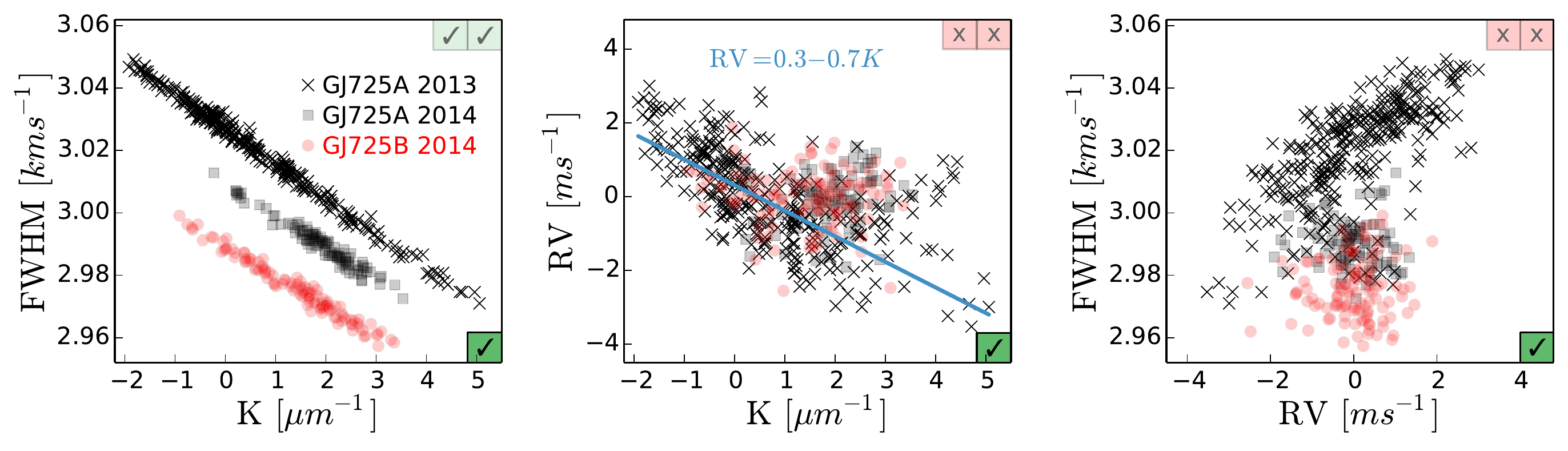}
	\caption{Spectral indices correlations for the five consecutive high-cadence nights of GJ~725A observed in August 2013 (black crosses). Alternating run observations taken in 2014 are also shown for comparison. The FAP is coded as in Figure~\ref{fig:corrimag} where the low right square refers to the 2013 data. The wider range of AM sampled in 2013 might have uncovered a chromatic effect also in the RVs (center). Blue line and equation correspond to the tentative linear fit we applied to correct the RV data (see Sec.~\ref{sec:detrending}). On the left plot, the negative slope of the 2013 FWHM$-K$ correlation is compatible with the 2014 results, pointing out towards an instrumental (or software) common origin. The FWHM$-$RV is shown in the right panel.} 
	\label{fig:corrspec13}
	\end{figure*}

As for the 2014 alternating run, the analysis shows the $K$-index clearly correlated
with the FWHM (see Figure \ref{fig:corrspec13}). During this run we covered a
wider range of airmasses -this study includes data only up to 2.5 in airmass-
compared with the 2014 alternating run (which covered up to 1.6 in airmass).
This might have been the cause of the larger span of $K$ values measured in
2013, but a punctual under-correction of the ADC can not be ruled out. The slope
of the correlation law is equivalent to the value measured on the differential
run. The offset could be due to a focus realignment carried out in between the
two runs (during the technical run of March 2014$^{\ref{fn:technical14}}$.) 

The wider range of variability in $K$ seems to better support the existence of a relation
between this index and the RVs as well (see Figure\ref{fig:corrspec13}, central plot). The
variability in both the FWHM and the RV suggested that the mean line profile was changing
during the night. However, we could not rule out a problem in the measurements themselves
(i.e. in the algorithms) producing spurious variability in the time-series. To address this, we
performed a number of tests and applied profile measurements independent from the DRS
in the next section.

%________________________________________________________________
\subsection{Independent measurements and validation experiments} \label{sec:experiment} 
%________________________________________________________________
Systematic variability in the measurements can be caused by \emph{intrinsic}
changes in the line profile (i.e. instrumental), by sub-optimal procedures in the measurement of
these quantities (i.e. algorithmic), or both. For example, even if the line-profile is perfectly
stable, inaccuracies in the fitting of the flux or background subtraction
will produce apparent variability of measurements of the line shape.
In particular, the cross-correlation profiles are obtained by
computing some weighted mean of the profiles of the individual spectral orders.
Because the effective line profile of each individual order is different,
changes in the weights (e.g. if the weights are computed using Poisson
statistics from the photon counts) will produce spurious variability correlated
with these changes. In other words, a different SED implies a different local SNR for each order. On the other hand, an inaccurate continuum or background subtraction will cause different mean line profiles at each spectral order. As a result, we can obtain a modified final mean line profile when we combine a variable weight distribution with different profiles at each order. Furthermore, the RV and FWHM values obtained from the final profile will be also different with the SED. Thus, if our methods do not perform a good continuum or background subtraction, we can measure spurious RV or FWHM variabilities even when the instrument is stable.

Such an effect was earlier reported by \cite{bourrier2014} on 55 Cancri. Following their findings, the DRS flux-normalises the continuum of all the spectra with respect to a reference spectrum for F, G and K stars. However this has not been yet implemented for M stars, because, as it was defined, some zero division problems can occur in the bluer part of the spectra (F. Pepe, private comm.)

In order to validate if the algorithms we used could be the cause of the observed variability in the measurements, we have performed the two experiments detailed below.    

%________________________________________________________________
\subsubsection{Experiment 1 : RV measurement against pSED variability}
HARPS-TERRA fits the flux iteratively. It carries out a least
squares fit to minimise the difference between the observations and a template
(built from the observations). In particular, the magnitude to minimise, $R$, is given by:

	\begin{equation}
		R(\lambda) = T[\alpha_{v}\lambda] - f[\lambda]\cdot \rm{FN},
	\end{equation}
	\begin{equation}
		\rm{FN} = \sum_{0}^{M}{\alpha_{m}(\lambda-\lambda_{c})^{m}},
	\end{equation}
where $T$ is the template, $f$ is the observed spectra, and FN the flux normalization term. Thus, HARPS-TERRA fits simultaneously the Doppler shift ($\alpha_{v}$) and the others parameters which account for the shape of the continuum ($\alpha_{m}$) at each order (m). As a consequence, the $\alpha_{m}$ parameters are also modeling instrumental flux variations effects such as the atmospheric differential refraction or tracking errors. 

In order to validate if this simultaneous continuum fitting technique applied by HARPS-TERRA prevents the RVs from being affected by chromatic correlations, we performed the experiment detailed below. 

\begin{figure}
	\centering
	\includegraphics[width=\columnwidth]{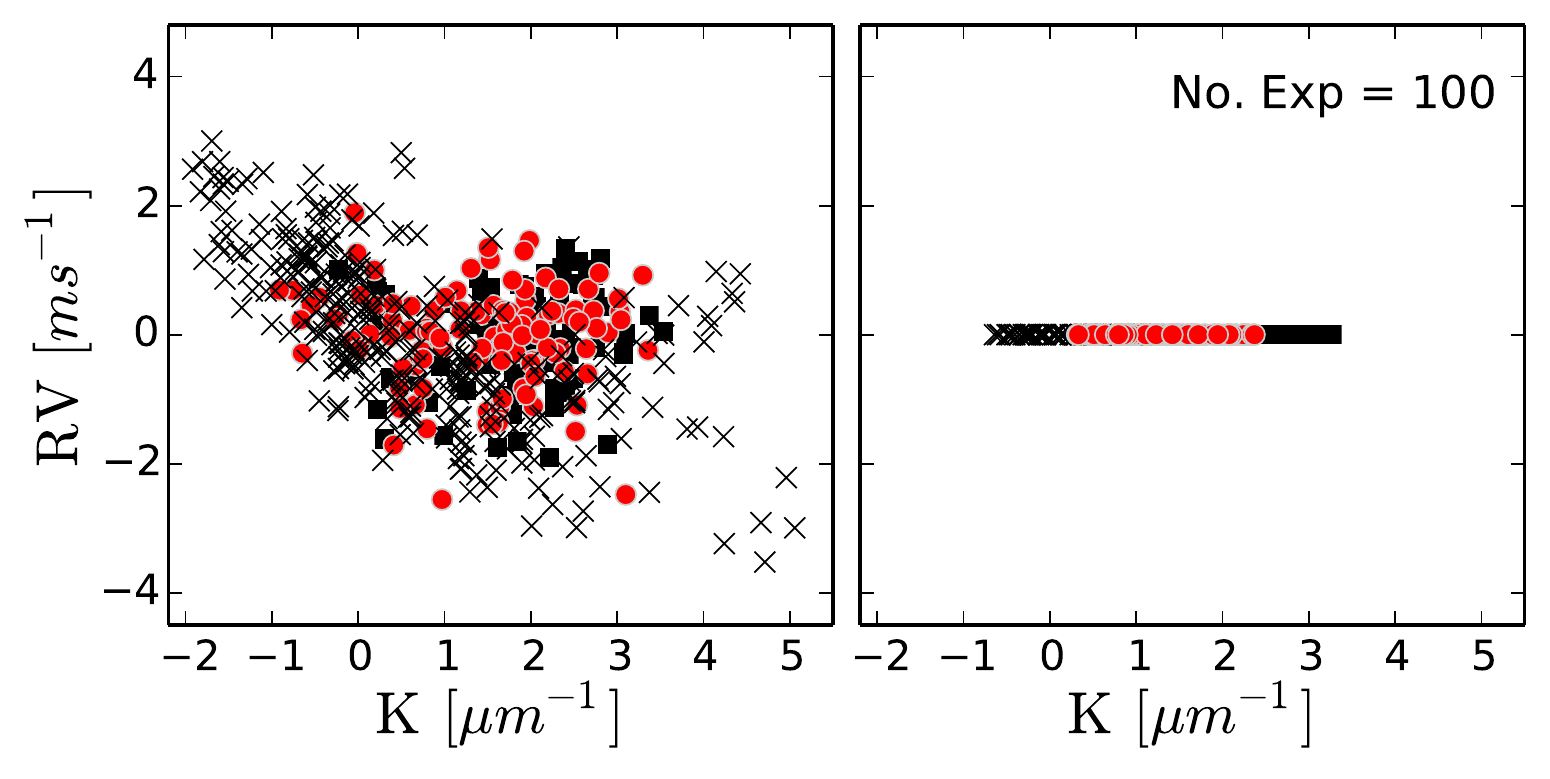}
	\caption{RVs plotted versus the $K$-index when the input corresponds to the real observations (left plot) or to a set of 100 flux-distorted spectra calculated from an observed epoch (right plot). HARPS-TERRA measures relatives values and centers the RV at $0~ms^{-1}$. The zero-values (RMS below $2\times10^{-6}~m\,s^{-1}$) on the right plot indicate that the normalization carried out by HARPS-TERRA ensures RVs independent on the flux distortions included deliberately to perform this experiment.}
	\label{fig:fluxexpht}
	\end{figure}

We selected three epochs, one for each star and run, and we changed the flux of their spectra as:
	\begin{equation}
		f_{\rm{new}}(\lambda, t) = f_{\rm{orig}}[\lambda, t] \cdot [1 + S(t) * (\lambda-\lambda_{\rm{cR}})],
	\label{eq:exp}
	\end{equation}
where $S(t)$ randomly takes values between 0 and 2.086 $\mu m^{-1}$. This range was selected to ensure positive new fluxes at least for the last 31 orders that were used in the second experiment (detailed in Section~\ref{sec:exp2}), and to result in flux distortions below 20\% (approximately the SED distortion measured in the linear range used to calculate $K$). We repeated this calculation several times, and, in total, we obtained 100 new synthetic spectra for each of the three selected epochs. Afterwards, we post-processed these synthetic spectra with HARPS-TERRA, recovering in all cases a RV with and RMS below $2\times10^{-6}~m\,s^{-1}$ for the three epochs (see Figure~\ref{fig:fluxexpht}).

This result indicates that the RVs measured with this template matching
algorithm are mostly insensitive to changes in the SED. HARPS-TERRA is properly
fitting the flux of the spectra, and thus, our intentionally introduced
wavelength distortions can not be the cause of the observed RVs variability
(neither of the RV$-K$ correlation).

%________________________________________________________________
\subsubsection{Experiment 2 : line profile width against pSED variability}\label{sec:exp2}

HARPS-TERRA provides an independent measurement of the RV, but we do not have
any equivalent for the FWHM. With the aim of checking if the strongest FWHM$-K$ correlation could be due to a software issue related with the weighting of the orders (see detailed explanation at the beginning of this section), we estimated the profiles using an independent least-squares deconvolution technique \citep[LSD;][]{donati1997}. In particular, we used the implementation given by \cite{barnes1998, barnes2012}. 

The least square routine involves solving the mean-line profile which, when optimally convolved with a line list, gives the best match to the observed spectrum. Generally, the line list used is just a theoretical identification of wavelengths and depths. Instead, we used the high SNR template computed by HARPS-TERRA to obtain a more realistic empirically determined list of lines. This template is built by co-adding all the observed 2-dimensional spectra given by the DRS (the so-called e2ds). Those regions of the spectrum with strong telluric absorption or strong stellar molecular bands were not used for the deconvolution.

Like the DRS, the LSD discards some wavelengths in the blue part of the spectra (only those corresponding to the last 31 orders were used, while the DRS uses the reddest 51 orders). However, in contrast to the DRS, the LSD does not calculate one profile per order. The deconvolution is performed on each spectrum using all lines (including those repeated in adjacent orders) to obtain one single LSD profile per exposure. Our LSD method normalises each spectrum to a template to obtain a individual continuum. That is, each spectrum in turn is divided by the template spectrum and a cubic order polynomial is fitted to the residuals. Later, the continuum of each spectrum is obtained by multiplying this cubic polynomial fit by the template continuum, which was previously obtained by iteratively fitting a 5th degree polynomial to the template. Because each spectrum is normalised to a individual continuum, we did not expect variability related with weighting problems arising from the algorithm or background issues (see beginning of this Section). 

The output of the LSD technique is a mean line profile in absorption. These $F_i(t)$ profiles can be inverted (sign changed) after have we subtracted their residual continuum $w(t)$, producing a normalised (positively defined) probability distribution function $f(t)$. Then,  the moments of $f(t)$ can then be computed as:
	\begin{eqnarray}
		f_i(t) &=& w(t)-F_i(t) \\
		\hat{M}_{n}(t) &=& \frac{ \sum_{i=1}^{N}{[f_{i}(t)]}\cdot v^{n}}{ \sum_{i=1}^{N}{f_{i}(t)}}
	\label{eq:moments}
	\end{eqnarray}

Before calculating the moments with Eq.~\ref{eq:moments}, we truncated the profiles in a range of $\pm~20~km s^{-1}$ around the estimated zero-velocity (the velocity of the co-added barycentric corrected high SNR template from which we obtained the line list applied during the deconvolution process), and we excluded the negatives values resulting after the continuum inversion. This truncation was done to avoid including wing distortions in the moments calculation. The profiles were calculated with a mean velocity increment of 0.81 $km s^{-1}$ --value that corresponds to the pixel size of the HARPS-N detector--, and later interpolated by a factor 10 using a spline function.  The order zero moment, $M_{0}$, corresponds to the integral of the normalized profile. $M_{1}$ is equivalent to its centroid  (all spectra are aligned to zero-velocity before computation of the LSD profile), and $M_{2}$ corresponds to the profile variance and thus, as the FWHM, is a measurement of the width of the line profile.

	\begin{figure}
	\centering
	\includegraphics[width=\columnwidth]{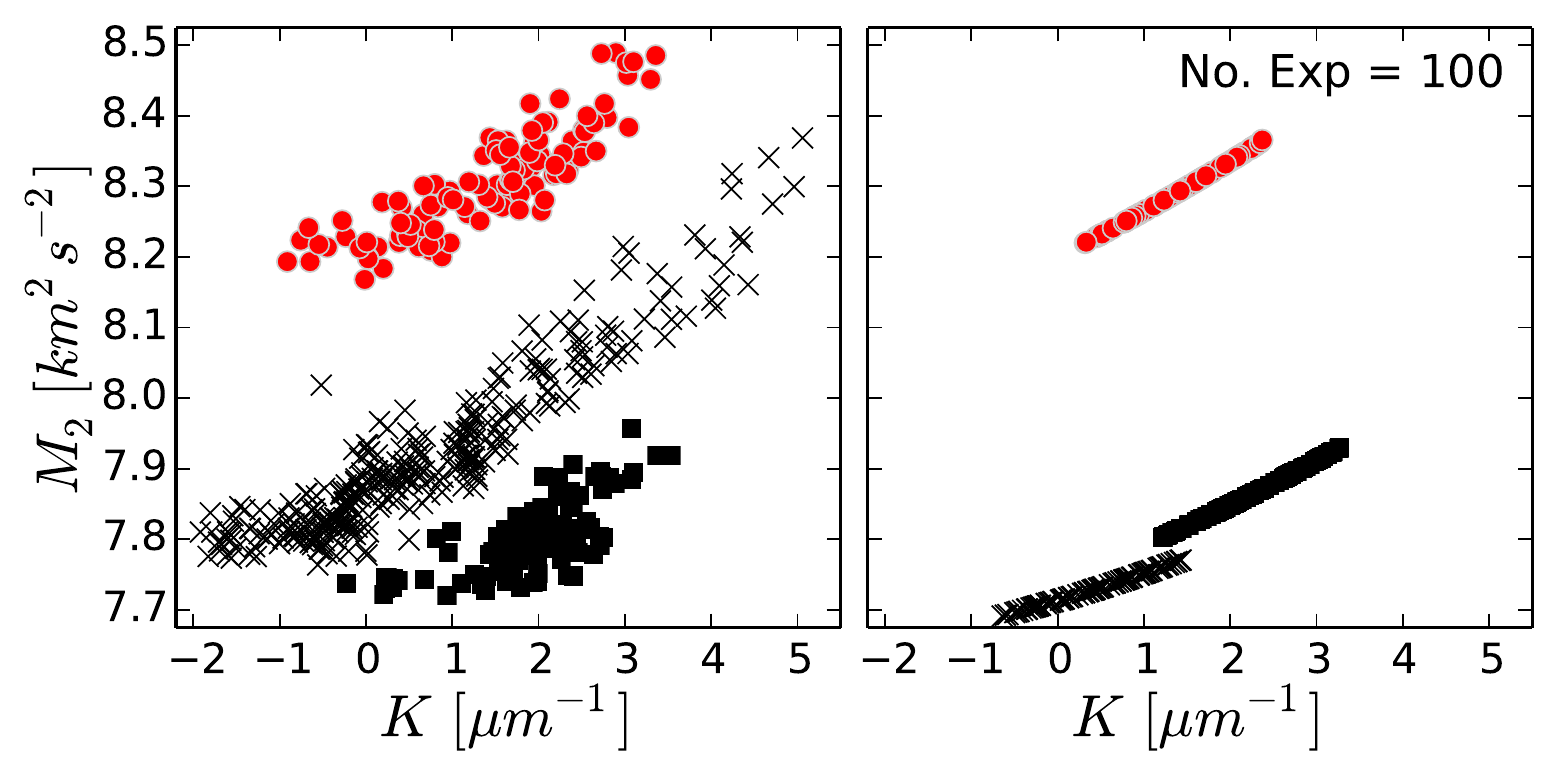}
	\caption{Moment of second order ($M_{2}$) of the LSD profiles versus the $K$-index for the observed (left plot), and for the synthetic (right plot) spectra. The trend for the synthetic data reproduces the observations pointing out a software issue.}
	\label{fig:fluxexplsd}
	\end{figure}

With the aim of testing the LSD performance, we calculated the LSD profiles of the flux
modified spectra that we used in the previous section to test the software HARPS-TERRA; i.e.
we took the same three spectra (one for each star and run) and we modified their flux
following the Equation~\ref{eq:exp}. In this case, the second order moment (or $M_2$)
calculated from these contaminated profiles, correlated with $K$ in a way analogous to the
observations (see Figure~\ref{fig:fluxexplsd}). This implies that, even with our method that fits the continuum of each spectrum individually, we find a strong correlation of the
measurement of the line width (through $M_2$) with the artificially introduced slope of the
pSED. In other words, the changes in the distribution of the flux along the CCD can drive the
FWHM variability. Then, the largest correlation involving the
FWHM$-K$ relation can be explained as an inaccurate computation of the profile by the
algorithms. Neither the DRS nor the LSD
method produce accurate enough measurements. Actually, the fact that we need to correct a residual continuum $w(t)$ for each deconvolution profile may arise from a non perfect description of the continuum. 

According to our experiment, only the RV measurements obtained by the HARPS-TERRA algorithm, seem to be insensitive to the SED variability. This result indicate that new algorithms that simultaneously fit for the continuum of each echelle order and the line profile need to be developed, or the current methods refined to account for this. 

To use the FWHM index produced with the M-dwarf mask of HARPS-N, one should -at the very least- apply some decorrelation procedure. Otherwise, all this
index is tracing are small changes in the measured SED produced by the expected flux loses on fibre-fed spectrographs.

%________________________________________________________________
\subsection{Detrending strategies for archival FWHM and RV obtained with HARPS}\label{sec:detrending}
%________________________________________________________________
Once we clarified that the strongest FWHM-$K$ correlation has an
algorithm origin, we aimed at defining an empirical function to detrend the FWHM
time-series from chromatic effects. The function should be valid for (at least)
the GJ~725A+B pair and other stars with similar spectral types (note that the
chromatic nature of the effect will likely affect different spectral types
differently). We made use of both, 2013 and 2014 data, to compute the detrending
law. 

The FWHM contains additional contributions from other effects. In
particular, since the observations span over a finite amount of
time, the FWHM will be broadened by the change in the barycentric
velocity of the observer. The following procedure was designed to remove 
the effect of this barycentric broadening (which can account for
several $\rm m\,s^{-1}$ depending on integration time and coordinates of the star).

The central moments of a distribution can be
denoted as $M_{k}=<(x_{i}-\bar{x})~^{k}>$. Thus, $\mathrm{M}_{2}$
= Var(x), the variance of the distribution which, for two
independent distributions, satisfies the additive property: Var(x +
y) = Var(x) + Var(x) \citep{booknumericalrecipes1992}. The
mean line profiles of the spectral orders can be considered as a
probability density function. In consequence, the FWHM should
also be a variance.  

\begin{figure*}
	\includegraphics[width=17cm]{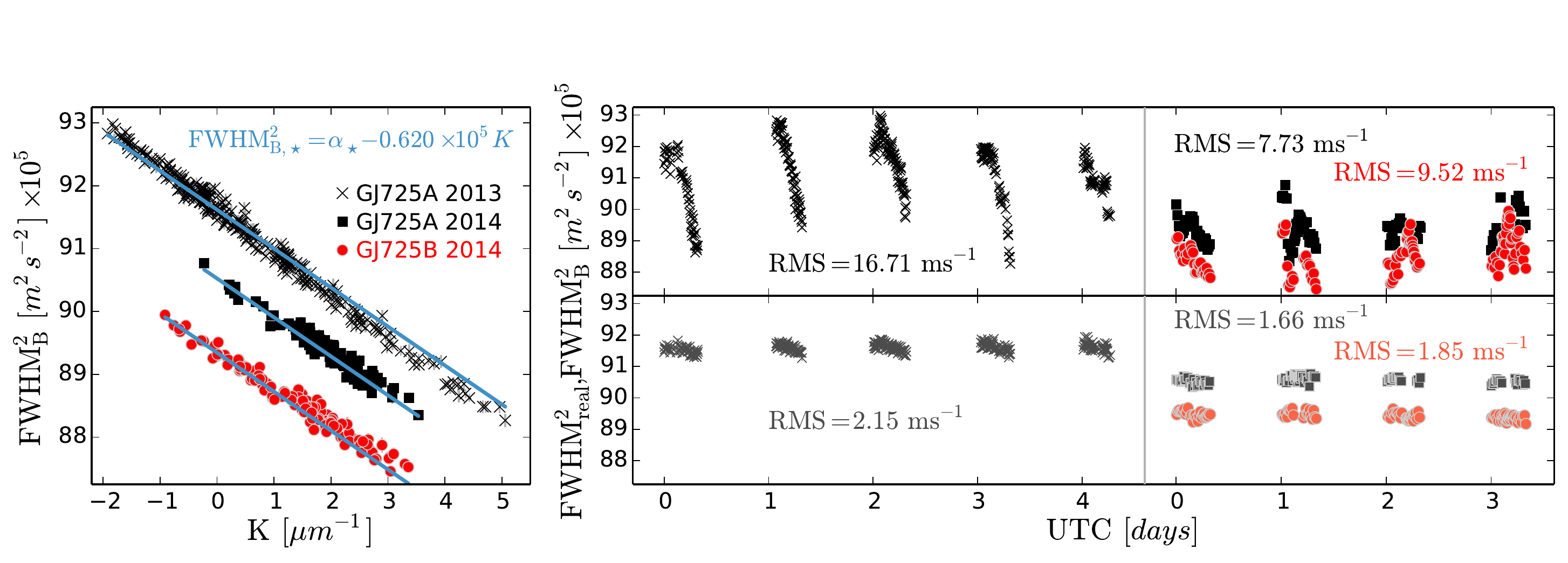}
	\caption{Left panel: FWHM (barycentric corrected) and $K$-index correlation. Black squares and red dots match GJ~725A and B, respectively. Blue lines and equation correspond to the best-fitting function. We want to emphasise that the empirical law is unique. Note that the chromatic correction consists of detrending the data by subtracting the slope term. Right panel: Time-series before ($\mathrm{FWHM}_{\rm B}^{2}$, top) and after ($\mathrm{FWHM}_{\rm real}^2$, bottom) the chromatic correction. The FWHM is flattened and corrected from chromatic distortions caused by the atmosphere. The RMS is reduced by a factor of 8 in 2013 and by a factor of 5 in 2014.}
	\label{fig:fwhmcorr}
	\end{figure*}

The observed FWHM$_{ \mathrm{obs}}^2$ is the sum of three terms:
	\begin{equation}
		\mathrm{FWHM}_{ \mathrm{obs}}^2 = \mathrm{FWHM}_{ \mathrm{real}}^{2} + \mathrm{B}^{2} + \mathrm{I}^{2},
	\end{equation}
where $\mathrm{FWHM_{\rm real}}$ is the true line profile we want to measure, and B and I are:
	\begin{itemize}
	\item{The barycentric correction (B):}
	It is a squared velocity value we have to subtract to the square of the observed FWHM to correct it from the movement of Earth on its orbit. To first order, let us assume that the velocity $v$ of the
observer changes between the initial instant $t_i$ and the
end-of-integration time $t_f$ by  $\mathrm{dV}$. The effect of the
barycentric broadening is then a convolution of the stellar profile with the double of the variance of a boxcar distribution ($F(v)$) of width $\mathrm{dV}$:
		\begin{equation}
		\mathrm{B}^{2}=2 \int_{-\infty}^{\infty}F(v)v^{2}dv = 2 \frac{\mathrm{dV}^{2}}{12},
		\end{equation}
where, being C a constant, $F(v)$ is the boxcar function $f(v)$ normalized to have unit area as follows, 
		\begin{equation}
		\,\,\,\,\,\,F(v) = \frac{1}{\mathrm{C\,dV}} f(v), 
		\end{equation}
		\begin{equation}
		\,\,\,\,\,\,f(v) = \left\{ 
		  \begin{array}{l l}
		   \mathrm{C} & \quad \text{if $\frac{-\mathrm{dV}}{2}<\mathrm{v}<\frac{\mathrm{dV}}{2}$}\\
		   0  & \quad \text{if $\mathrm{v}<\frac{-\mathrm{dV}}{2}$ or $\mathrm{v}>\frac{\mathrm{dV}}{2}$}
		  \end{array}, \right.
		\end{equation}
This correction can be exactly computed using the observation,
the exposure time, and a custom made code used to compute the barycentric correction
(implemented within the HARPS-TERRA software).
\\

	\item{Chromatic effect correction (I):} 
This term contains the correlation with the chromatic $K$
index as computed in previous sections. We searched for the linear
model which best-fitted the observed FWHM corrected from barycentric term ($\mathrm{FWHM}_\mathrm{B}$), and the $K$-index series,
		\begin{equation}
		\mathrm{FWHM}_{\rm B}^{2}= \mathrm{FWHM}_{\rm obs}^{2}- \rm B^{2}=\alpha + \beta\; K,
		\end{equation}	
where  $\alpha$, the offset, is a nuisance parameter, which is equal
to $(90.53\pm0.02)\times10^{5}\rm\,m^{2}s^{-2}$ for GJ~725A, 
$(89.34\pm0.01)\times10^{5}\rm\,m^{2}s^{-2}$ for GJ~725B, and to
$(91.620\pm0.004)\times10^{5}\rm\,m^{2}s^{-2}$ for GJ~725A observed in the 2013. 

Thus, we defined the chromatic correction term as
		\begin{equation}
		\mathrm{I}^{2} = \beta\;K,
		\end{equation}	
where $\beta = (- 0.620\pm0.003)\times10^{5}\,\rm m^{2}s^{-2}\mu m$
is the average of the values obtained for the three data sets
independently. The low degree of scatter in the values (coefficient of
variation lower than 10 per cent) means that the average is
representative and thus, we can define a common law which is the
same for both stars and all observed nights (see
Figure~\ref{fig:fwhmcorr}, left panel).

	\end{itemize}
		
Finally, the FWHM corrected from barycentric and chromatic effects can be written as:
	\begin{equation}
	\mathrm{FWHM}_{\rm real}^2 = \mathrm{FWHM}^{2} - 2\,\frac{dV^2}{12} - \beta \;K,
	\end{equation}
and its error, obtained by applying simple error propagation functions, as:
	\begin{equation}
	\sigma_{\mathrm{FWHM}_{\rm real}^2} =\sqrt{ 4\; \mathrm{FWHM}^{2} \; \sigma^2_{\mathrm{FWHM}} +  K^2 \sigma^2_{\beta}  }
	\end{equation}

Results are shown in Figure~\ref{fig:fwhmcorr}. The same slope for 2013 and 2014 data was expected as it is caused by the same instrumental effect. However, results indicated some systematic residual effects, specially in the 2013 data. This extra variability might be related to some lesser degree contributions specific to each run. However, as the main correlation might have the same origin, we still prefer to apply the same empirical law to detrend all the datasets. We also expected the zero-point offset ($\alpha$) of the two runs to be the same for GJ~725A, but the change could be due to an instrument focus readjustment carry out between the two observational campaigns. In spite of these discrepancies, the improvement is
obvious, the RMS is reduced a factor of 5: GJ~725A from 7.73 to
$1.66\rm\,m\,s^{-1}$, GJ~725B from 9.52 down to
$1.85\rm\,m\,s^{-1}$; and a 87 per cent for the 2013 run in GJ~725A, from 16.71 down to $2.15\rm\,m\,s^{-1}$.  We want to point out that the reduction of RMS comes from the presence of highly structured noise,
which makes the new time-series of the FWHM a much more reliable
tracer of the physics occurring in the photosphere of the star.

On the other hand, the intranight systematic in the RVs can be also modeled with a linear
function. \cite{bourrier2014} detected similar RV correlation for
the 55~Cnc high-cadence data observed with HARPS-N. They used the
ratio of SNR of two orders as the detrending quantity, which is a
simpler version of our chromatic $K$ index. As discussed before, we did not find any significant correlation for the 2014 run. 
Therefore, we have only detrended the 2013 data, following:
		\begin{equation}
		\rm RV_{\rm C} = RV - \omega\,K,
		\end{equation}
where $\omega = -0.7\pm0.2\;\rm m\,s^{-1}\mu m^{-1}$. This value of $\omega$ is obtained by fitting the linear correlation law to each night independently to avoid
contamination by additional, longer term variability (e.g. planets or
induced activity happening at time-scales of a few days) Finally, the fitting parameters (see equation in the central panel of Figure~\ref{fig:corrspec}) are the weighted mean of the i-th values.

	\begin{figure}
	\centering
	\includegraphics[width=\columnwidth]{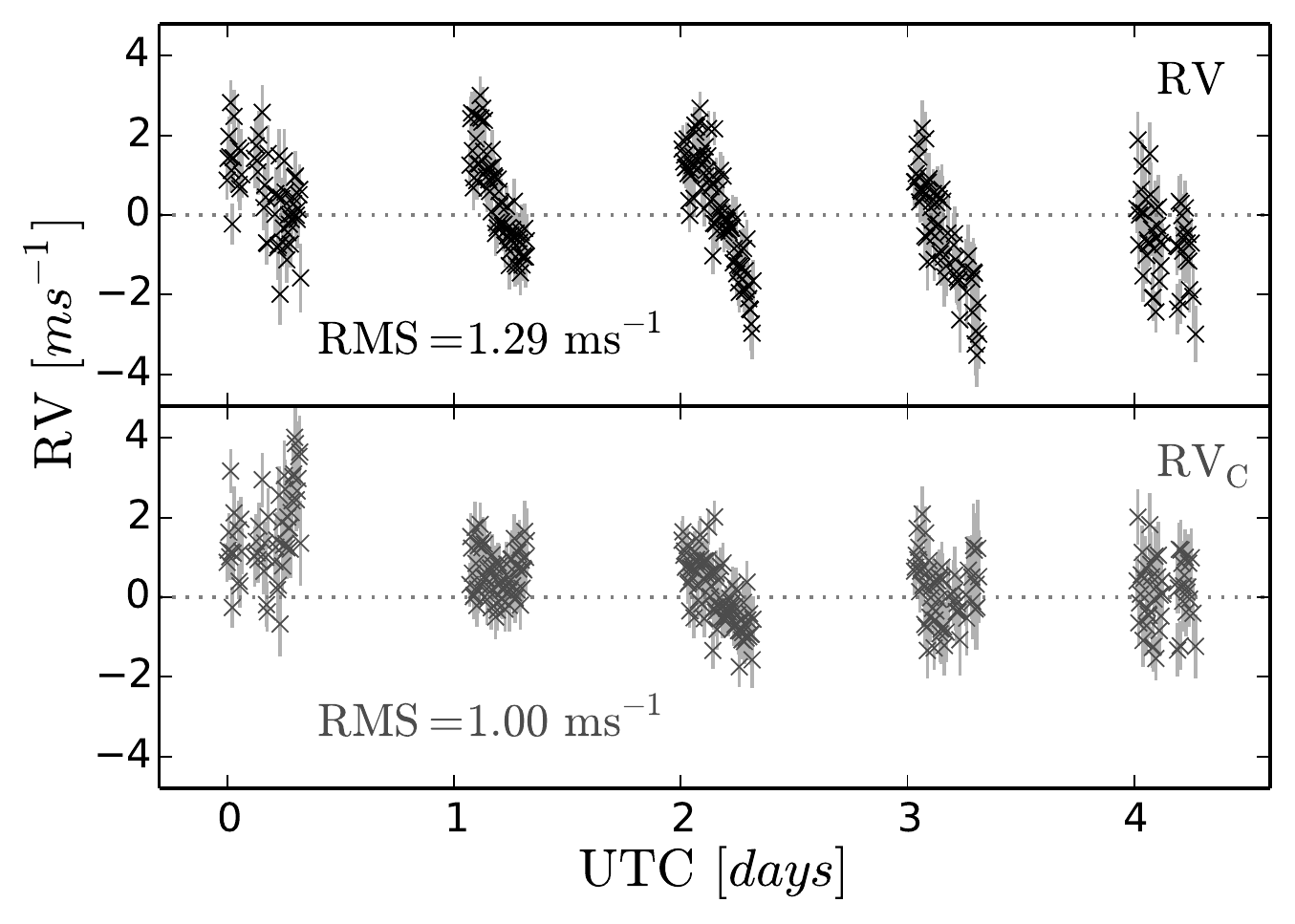}
	\caption{2013 RV time series of GJ~725A before (top panel) and after (bottom panel) applying the empirical function to detrend the RV-$K$ correlation. The RMS decreases by a factor 1.3.}
	\label{fig:rvcorr}
	\end{figure}

Detrended RVs, shown in the bottom panel of Figure~\ref{fig:rvcorr},
have a mean peak-to-peak difference of 3.3~$\rm m\,s^{-1}$. The RMS
is reduced from 1.29 down to 1.00~$\rm m\,s^{-1}$, which is close to
the photon noise of the HARPS-N observations. 

We advise caution in using the correlation laws defined here to detrend the FWHM and RVs from HARPS-N observations. As we have seen, instrumental updates and (possibly) spectral
types might produce slightly different values for the correlation laws. As a general rule, in case of aiming at detrending archival data, firstly we strongly recommend to compute the chromatic $K$ index, and secondly, verify if there are chromatic effects; that is, verify if there are correlations between the $K$ index and the main data-products (e.g. RVs, FWHM or other indices). Finally, in case of detecting significant correlations, follow the detrending steps explained in this study. We want to note that, in case of detecting correlations with the RVs, it is preferable to include $K$ as a correlation term in the model used to search for Keplerian signals instead of applying pre-whitening like methods \citep[see an example in][]{anglada2014}. 

%________________________________________________________________
\subsection{Cross-dispersion profiles} \label{sec:spaprof} 
%________________________________________________________________

The cross-dispersion profile, or the mean profile obtained from making a cut of the \'echelle
spectrum across the orders, is very sensitive to illumination distortions produced at the
fiber entrance such as changes of the telescope focus, seeing increases or pointing errors.
The reason for this is that each cut across the order is a section of the fiber image; in
fact the fiber size defines the width of the order. Therefore, the cross-dispersion profile
is by itself a measure of the instrumental profile and can be used to measure the
illumination stability in the detector. For example, if the wavelength dependence variations
of the flux of the spectra were due to a non-perfect smoothness of the instrumental profile
(e.i. due to an inefficient fiber scrambling system on HARPS-N; \citealp{avila2004}), we
would expect to find a correlation between the width of the cross-dispersion profile and the
$K$ index.
	\begin{figure}
	\centering
	\includegraphics[width=\columnwidth]{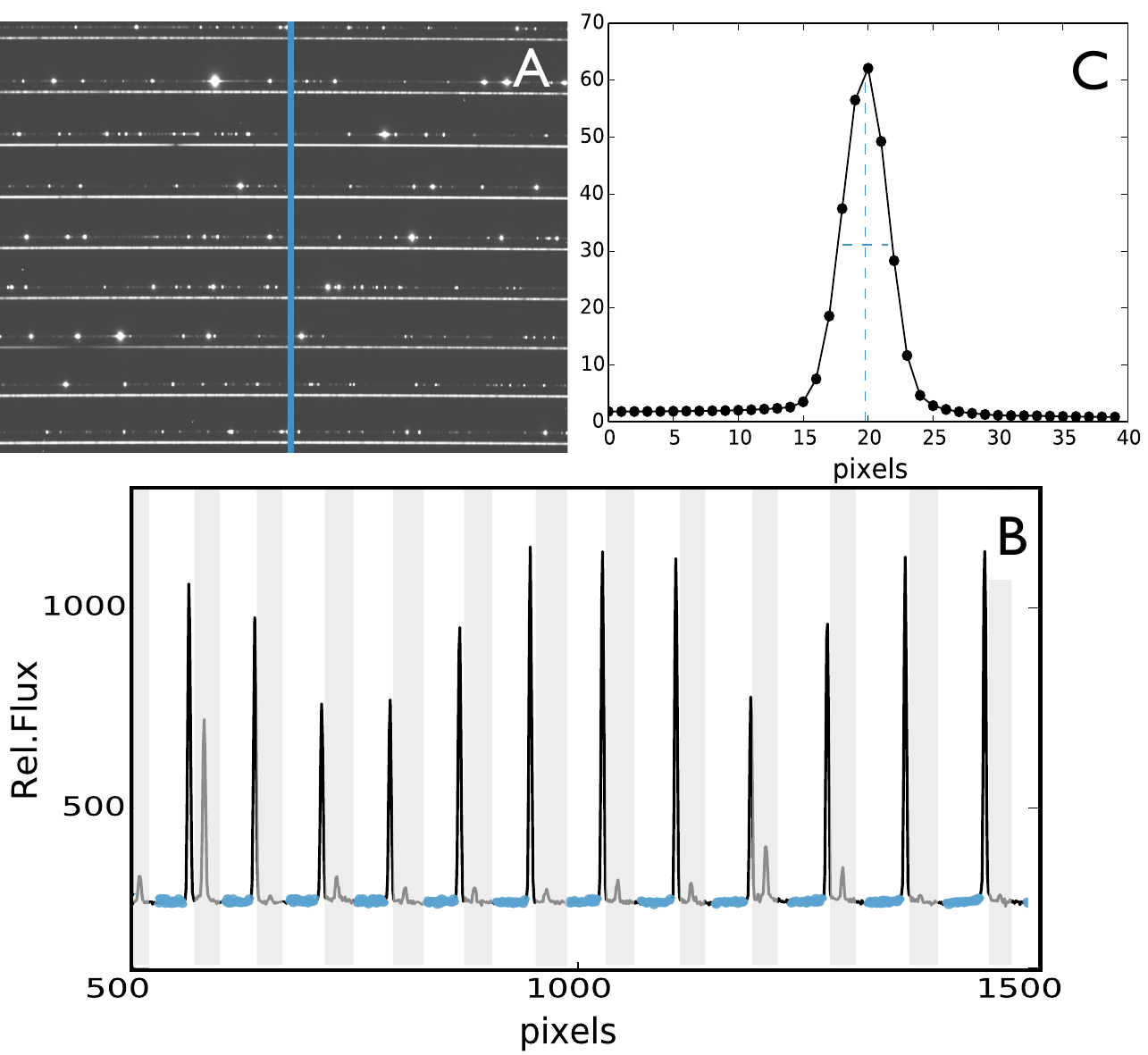}
	\caption{Layout of the cross-dispersed profile calculation process. To extract the cross-dispersed profiles (B), the raw images are cut in the spatial axis following the direction of the blue line (A). Simultaneously with the science observations, we recorded emission lines from a ThAr lamp (see bright spots in the raw image). We blocked the lamp lines (B, shaded areas) before selecting the points used to fit and subtract the floor level (B, blue dots). Finally, after normalizing each peak to its maximum, we cross-correlated the profile with the flat raw image to obtain a mean line profile of the column (C).} 
	\label{fig:digram}
	\end{figure}

To measure this profile, we cut the raw images across the cross-dispersed axis, extracting in that way the flux at each column of the reddest half of the CCD (see diagram in Figure~\ref{fig:digram}). After blocking the lamp emission, which is recorded in the science image simultaneously using a secondary fiber, we fitted and subtracted the floor level using a polynomial. The counts obtained in between consecutive science orders were compatible with the bias level obtained during the standard calibration performed at the beginning of the night. However,  we fitted a polynomial instead of subtracting a constant bias level to better account for other possible background effects that can introduce biases in the final width of the cross-dispersion profile. Then, we normalised each order to its maximum and we cross-correlated each column of the raw image with the corresponding flat column. After that, we fitted a Gaussian to this cross-correlated profile, calculating the cross-dispersed FWHM at each column as:
	\begin{equation}
		\mathrm{FWHM} = 2\sigma\sqrt{2\log{2}}, 
	\end{equation}
where $\sigma$ corresponds to the width of the fitted Gaussian. We note that this function includes a fourth parameter (an offset) in an attempt to model the background flux variabilities. 

Because the detector is flat, the image of the fiber onto the detector --and thus the FWHM in the cross-dispersed direction-- increases towards the detector edges. To avoid this effect, we took the differential cross-dispersed FWHM measurements of each column with respect to the first epoch. Furthermore, the detector edges are more sensitive to illumination changes (see table~7 of the HARPS-N User Manual\footnote{http://www.tng.iac.es/instruments/harps/}), and this cause the measurements at the edge columns to be more sensitive to SNR variabilities. Aiming at correcting the SNR dependencies of this origin, we detrended the cross-dispersed FWHM using the fitted offset parameter from the Gaussian model. The final measurement for each exposure ($\rm FWHM_{\rm cd}$) was obtained as the median along the column detrended values. 

We have also calculated the centroids in the cross-dispersed axis. This measurement corresponds to the mean parameter of the Gaussian function fitted to the profile. The cross-dispersed centroid (or $\rm CEN_{\rm cd}$) of the exposure was directly defined as the median of all the column values. In this case, we could not use the background level to detrend this measurements because, as it was expected, the centroid of the Gaussian did not move with changes of the background flux level; as a consequence we still had some correlation of the $\rm CEN_{\rm cd}$ with the SNR.

	\begin{figure}
	\centering
	\includegraphics[width=\columnwidth]{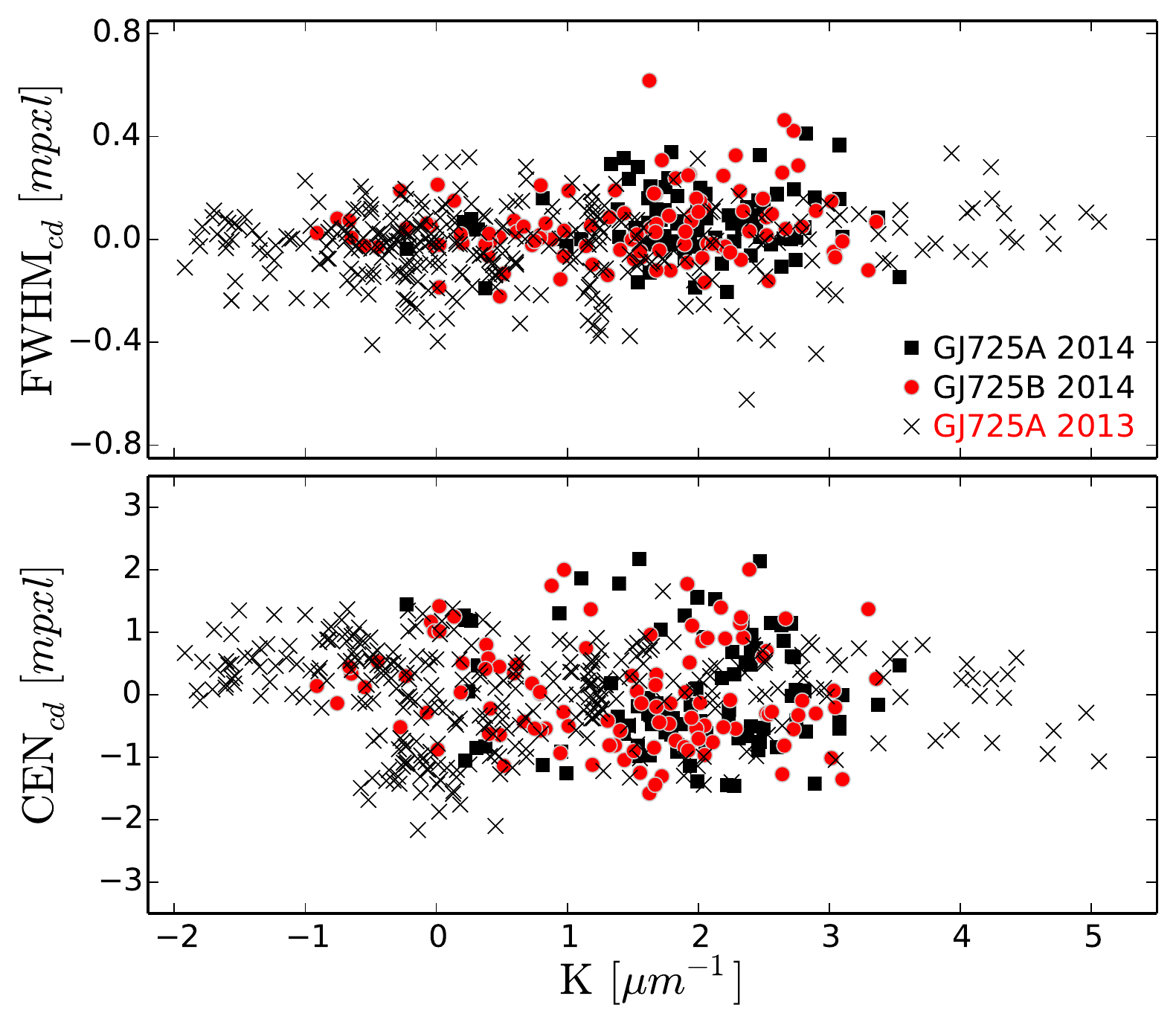}
	\caption{FWHM (upper panel) and centroids (lower panel) fitted to the mean line profile calculated on the cross-dispersed direction (perpendicular to the spectral axis). No significant correlation with the $K$ index was measured, ruling out a scrambling issue. The $\rm FWHM_{\rm cd}$ values are referred to the first exposure of the series. The average values of the three $\rm CEN_{\rm cd}$ series were subtracted to better compare its scattering.} 
	\label{fig:spaprof}
	\end{figure}

In spite of the corrections explained above, we detected the $\rm FWHM_{\rm cd}$ to be more
dispersed at low SNRs. Therefore, we performed an additional experiment to discard possible
non-linearity effects of the detector as a source of variability. We let the values of the cross-correlated profiles
of the 20\% of the exposures with worse SNR to randomly vary $\pm~0.2$~\%, total amount that could be associated to a non-linearity effect\footnote{See the ESO report
about the EEV44-82 CCDs at:  https://www.eso.org/sci/facilities/develop/detectors/optdet/
docs/reports/EEV-report.html}. However, we found no significant changes, discarding the non-linearity as the responsible for any change in the $\rm FWHM_{\rm cd}$.

If the spectral FWHM variability was due to illumination effects (instead of being a pure
software issue), the variation of the cross-dispersed FWHM with $K$ would be significant.
However, as we show in Figure~\ref{fig:spaprof}, the scatter of the $\rm FWHM_{\rm cd}$ is
below $1~mpxl$ ($1~mpxl\sim0.8~m\,s^{-1}$). Moreover, we did not detect any significant
correlation with $K$. The cross-dispersed centroids, whose maximum RMS was only $0.8~m\,s^{-1}$, do not correlate with K either, opposite to what we detected for its spectral counterpart in the 2013 observations.

In consequence, we do not detect significant distortions of the image onto the detector through the measurements in the cross-dispersed direction. Bearing in mind that, in the case of the $\rm CEN_{\rm cd}$ we still deal with some correlation with the SNR, this result indicates that, neither the large variability detected on the spectral FWHM nor the RV-$K$ correlation measured for the 2013 data, are due to a non-perfect smoothness of the inhomogeneities of the injected light spot. This result validates the use of the cross-dispersed profiles as a useful test to account for
illumination issues.

%________________________________________________________________
\subsection{Possible GJ~725B planet candidate?} \label{sec:dopplerb}
%________________________________________________________________
In this section we discuss a possible planet hosted by GJ~725B. As we pointed out in Figure~\ref{fig:difrunper}, the F-ratio excess detected in the RV differential periodogram could be compatible
with a planet orbiting either GJ~725B or GJ~725A.

To assess the significance of the variability and its
possible fit to a Keplerian orbit, we use likelihood
periodograms as described in \citet{baluev2009} and in \citet{anglada2013}. The model:
\begin{equation}
v_{r}(t) = \gamma+A\cos\left(\frac{2\pi}{P}\right)\Delta t+B\sin\left(\frac{2\pi}{P}\right)\Delta t,
\end{equation}
includes two sinusoids (2
parameters, A and B) plus and velocity offset (+1 parameter, $\gamma$). We used this circular model after have obtained an eccentricity compatible with zero with the complete Keplerian model. Like in classic periodograms, the likelihood periodograms produce a map of peaks for each of the investigated periods. However, in this case the power of the peaks is defined by the model that maximises the logarithmic of the likelihood statistic (\L). The \L\ maximisation and $\chi^2$
minimisation are equivalent, but the likelihood
method has the advantage, with respect to
traditional periodograms, of including the so-called jitter as a
free parameter (excess white-noise, +1
parameter). Using the likelihood statistic has a
further advantage: \L\ gives a probability, and the
ratio of likelihoods $\Delta$\L$\,=
\mathcal{L}_{\mathrm{null}}-\mathcal{L}_{\mathrm{model}}$
directly estimates the relative probabilities between
models. Properties of the likelihood periodograms are
described in full detail in \cite{baluev2009,
baluev2012}.

The likelihood periodogram analysis associates the peak
detected in the RVs differential periodogram with a $2.7\pm
0.3$~day signal with an amplitude of $1.2\,\rm m\,s^{-1}$. This
means that a model of a sinusoid at that period has a
$\Delta$\L = 32.79  when compared with a model with no signal.
This corresponds to a false alarm probability (FAP) $<
1.5\times 10^{-11}$, which is much smaller than the usual 1\%
threshold. A compatible period is obtained when analyzing
GJ~725B RVs independently, indicating that the variability
comes from this star. Besides the common variability in both
stars, no additional significant RV variability is detected on
GJ~725A.

	\begin{figure}
	\centering
	\includegraphics[width=\columnwidth]{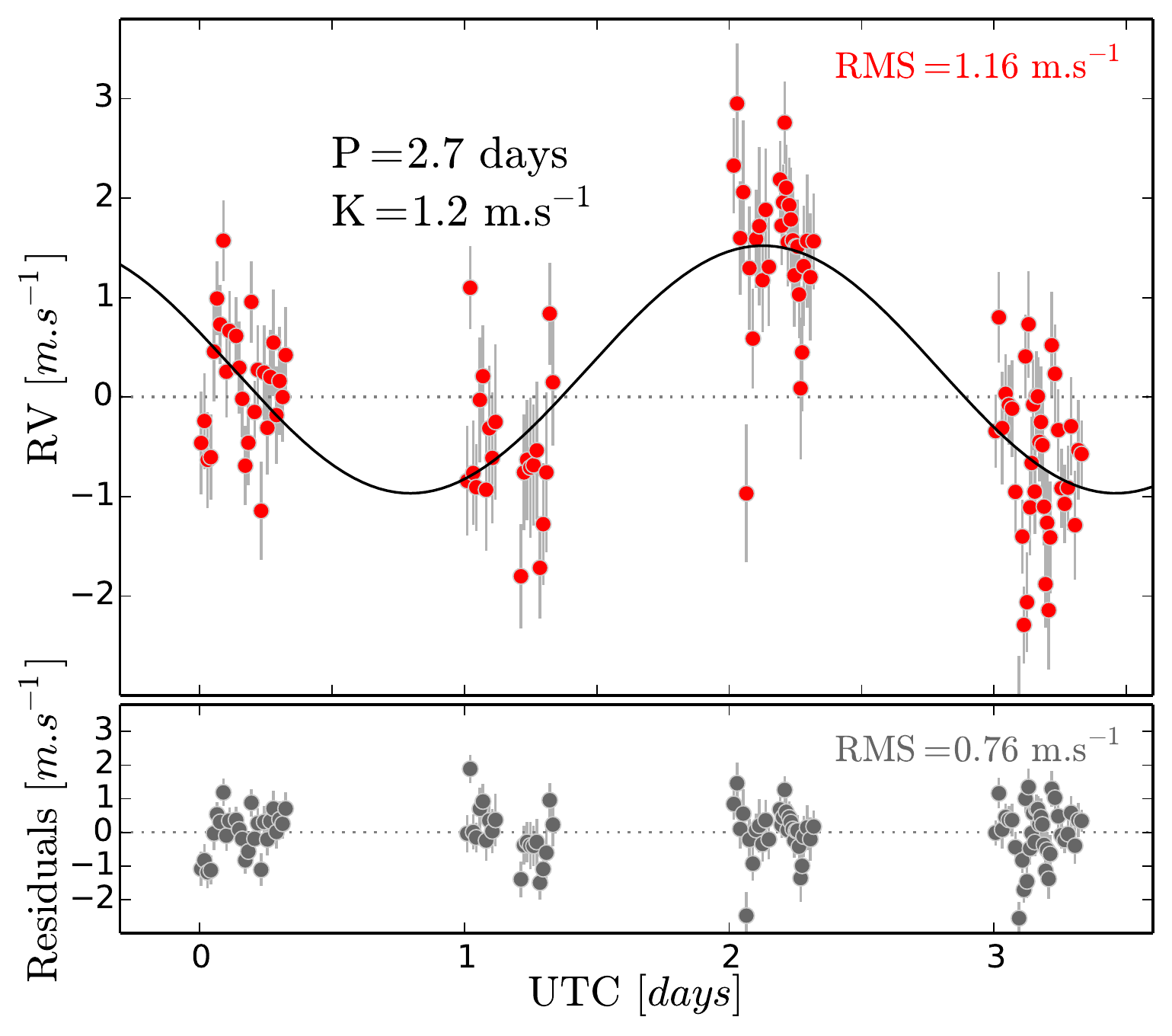}
	\caption{GJ~725B planet candidate signal. Red dots in the top panel are the GJ~725B RVs for the 2014 run. The black line is the circular Keplerian model best-fitting the data (P~=~2.7~d, $\rm K~=~1.2~\rm m\,s^{-1}$, $\rm FAP~<~0.1\%$ and $\Delta$\L$~=~49.13$). Residuals to this model are plotted in the lower panel. The RMS decreases from 1.16 to 0.76~$\rm m\,s^{-1}$ on the residuals. Time is referenced to the first exposure of the run.}
	\label{fig:doppler}
	\end{figure}

The variability in GJ~725B is consistent with a 1.2~\mearth\
planet with an orbital period of $2.7\pm0.3$~d period (or
0.025 AU, see Figure~\ref{fig:doppler}). While the statistical
significance is very high (a sinusoidal model is massively
preferred over no signal at all), its nature cannot be
confirmed because we cannot verify its strict periodicity.
Doppler follow-up observations and/or transit searches will be
needed to confirm it.

%________________________________________________________________
\section{Discussion and Conclusions} \label{sec:conclusions}
%________________________________________________________________

We obtained high-cadence observations using the HARPS-N spectrograph
on the nearby M-dwarf binary system GJ~725A+B. The two stars alternating
observations unveiled common strong systematic effects in the measurements 
of the width of the mean line profile (through the FWHM), and also in the RV
measurements to a lesser degree. The presence of these systematic
effects is a likely component of the floor noise observed in long-term Doppler programs of M-dwarf stars, and seriously affects the use of high-cadence observations for very low amplitude signal searches ($<2\,\rm m\,s^{-1}$).

The systematic effects seem to be related to flux loses due to imperfect corrections of
chromatic effects introduced by the Earth's atmosphere at the telescope-fibre interface. This
suspicion is motivated by several measurements and correlations observed in pre-fibre images
of the star compared to a number of measurements on the spectrum, most notably changes in the
slope of the measured SED. The comparison of measurements between the two stars (and their
almost identical systematic behaviour) shows beyond reasonable doubt that the significant
variability observed with high-cadence on M-dwarfs has an instrumental origin. While
most variability in the width of the line profile is likely to be caused by the algorithms
used, smaller residual RV variability remains unexplained.

HARPS-N includes an ADC which corrects the atmospheric dispersion; however other sources of
error such as the atmospheric extinction or the wavelength dependence with the seeing remain
uncorrected. A direct consequence of a non-optimal correction of the Earth's atmosphere is a
superposition of wavelength-dispersed images of the star at the fiber entrance. This causes
the light injected into the fiber to vary in wavelength as the energy peak (used for
centering the image in the fibre during exposure) changes with the airmass. To measure
whether this was causing distortions in the spectra, we defined the $K$ index, which measures
how the flux is distributed across the detector during the observations (i.e. accounts for
the SED variability). Besides, we measured variability of the $K$ index. This, as well as
with the airmass, resulted to be correlated with the distortions on the pre-fiber images.
However, while these SED variabilities cause changes in the flux, this does not necessarily
imply that the instrumental profile shape changes too (neither the data-products) due to
illumination effects. In fact, to avoid illumination dependencies, all current
high-resolution fiber-fed spectrographs apply scrambling methods \citep{avila2004}.
Therefore, a correlation of the SED variability with the data-products is not expected.
Nevertheless, we found a strong correlation of the SED variability with the line-width (through
the FWHM) and a correlation with the RVs only for our 2013 single star run on GJ~725A.
Further measurements of the spectral order shapes in the cross-dispersed direction
ruled out an inefficient scrambling as the origin for these correlations. These measurements
also validate the use of the cross-dispersed profiles as a useful test to account for
illumination issues on other echelle spectrographs.

We performed a couple of test to validate the algorithms used in this work. Results indicate that, whereas HARPS-TERRA properly corrects and models the continuum, the DRS is sensitive to changes in the slope of the SED. This computational issue explains the strong FWHM-$K$ correlation, however, 2013 RV-$K$ correlation remains unexplained and has to be caused by another effect. Since the width of the mean line profile is a parameter very sensitive to magnetic activity events \citep[e.g.][]{reiners2013}, the FWHM is often used as an activity indicator and a tool to decorrelate Doppler time-series. Thus, there is a value in detrending this index. We outline a procedure to decorrelate the HARPS-N FWHM measurements using the slope of the SED. A tentative decorrelation law is also proposed for the 2013 RVs. The decorrelation laws are likely to be slightly different on each target (especially for targets of different spectral types), so we advise to obtain the SED slopes changes with the $K$ index and perform similar verifications on all individual targets of a given programme.

Ideally, new algorithms that simultaneously fit for the continuum and the line profile need to be developed to account for the FWHM variability. Other option is to refine the current methods, as it was already done in the last DRS version for F, G and K stars, where the continuum is re-normalised with respect to a reference spectrum. Regarding the
variability of the RVs in the 2013 single run in GJ~725A, given the large
correlation with the airmass and the erroneous updating of the ADC movement pointed out during other campaigns, we suspect that the most likely explanation for it was the
sub-optimal performance of the atmospheric correction. The possibility of a systematic ADC
failure only during continuous mode observations (where the pointing procedure is not redone between
same object exposures) was ruled out by the HARPS-N core team (R. Cosentino, priv. comm.). However, our in-situ monitoring of the ADC parameters during posterior HARPS-N campaigns showed that, under some unknown circumstances, the ADC values seem no to be updated. Therefore, we suspect that the ADC failed during this 2013 run.

Remaining correlations with the image distortions might be the origin for the noise floor level which popped up as common signals in the GJ~725A+B alternating run. As a by-product, we have shown that the GJ~725A+B pair is stable enough to be used as a benchmark case for commissioning of future high precision spectrographs, at least down to $\sim1-2\,\rm m\,s^{-1}$. Indeed, CARMENES spectrograph \citep{quirrenbach2014} is planning to use this system for validation purposes. The $\sim1-2\,\rm m\,s^{-1}$ limit is possibly set by a low-mass companion orbiting GJ~725B, whose presence will be further investigated in future campaigns.

Long-term surveys (like the HARPS-GTO program), try to observe all the stars at the same
airmass each night, minimizing sources effects related to airmass and scrambling, and
randomizing possible systematic shifts occurring within a night. However, this is not
possible for programs which require continuous high-cadence observations. Many science
cases, such as the molecule detection on transiting exoplanets
\citep{snellen2010, martins2013}, can benefit from an extremely stabilized instrumental
profile and robust procedures to measure it. Concerning Doppler spectroscopy, we note that an
extra effort should be put into the image stabilization and correction of chromatic effects
such as those reported here ($\sim1\,\rm m\,s^{-1}$) in order to clean spurious signals and
get the most of the new technologies for wavelength calibration at $\rm cm\,s^{-1}$ levels,
like laser frequency combs \citep{probst2014}. The best wavelength determination is useless
if systematic signals populate the periodograms.

%-----------------------------------------------------------------------------------------------------------
% ADDITIONAL CONTENT
%-----------------------------------------------------------------------------------------------------------

\begin{table*}
\begin{center}
\caption{Radial Velocities observed for GJ~725A and GJ~725B with HARPS-N. The columns are: barycentric Julian date, RVs calculated with HARPS-TERRA and their errorbars, FWHM of the cross-correlation function, signal-to-noise at the spectral order centered at 631 nm, airmass,  the $\Phi$ and $\Theta$ indices obtained from the autoguide camera images, the $K$-index measured over the spectra, the FWHM and the RV corrected from intranight systematics.}     
\label{tab:data}
\begin{tabular}{lcccccccccc}\hline
OBJECT  &BJD & RV & FWHM & SNR & AM & $\Phi$ & $\Theta$ & $K$ & $\mathrm{FWHM}^{2}_\mathrm{real}$ & $\mathrm{RV_{C}}$  \\
                &  & $\rm m\,s^{-1}$ & $\rm m\,s^{-1}$ & at 631~nm& arcseconds & pxl & arcseconds & $\mathrm{\mu m}^{-1}$ & $(\rm{\rm m\,s^{-1}})^{2}$& $\rm m\,s^{-1}$ \\ \hline         
GJ~725A & 2456518.36 & 0.88 & 3024.13 & 66.48 & 1.22 &   ---    &   ---    & 0.00 & 3024.13 & 0.88 \\
                & 2456518.37 & 1.42 & 3030.39 & 72.79 & 1.22 &   ---    &   ---    & -0.34 & 3026.87 & 1.18 \\
                & 2456518.37 & 1.98 & 3030.63 & 63.74 & 1.21 &   ---    &   ---    & -0.50 & 3025.55 & 1.63 \\
                & ... & ... & ... & ... & ... & ... & ... & ... & ... &\\
                & 2456841.39 & -0.71 & 3002.59 & 67.28 & 1.55 & 6.85 & -0.04 & 0.68 & 3009.60 & ---\\
                & 2456841.40 & -0.86 & 2996.76 & 65.87 & 1.48 & 6.93 &  0.00 & 1.26 & 3009.73 & ---\\
                & 2456841.41 & -0.36 & 2991.07 & 53.23 & 1.42 & 7.85 &  0.00 & 1.76 & 3009.26 &  ---\\
                & ... & ... & ... & ... & ... & ... & ... & ... & ... &\\
GJ~725B & 2456841.39 & -0.46 & 2984.25 & 52.24 & 1.51 & 7.24 &  0.01  & 0.66 & 2991.11 & ---\\
                & 2456841.40 & -0.24 & 2985.22 & 59.62 & 1.45 & 6.70 & -0.05  & 0.75 & 2993.00 & ---\\
                & 2456841.42 & -0.63 & 2978.01 & 58.27 & 1.39 & 6.58 &  0.09  &  1.50 & 2993.54 & ---\\
                & ... & ... & ... & ... & ... & ... & ... & ... & ... &\\ \hline
\end{tabular}   
\end{center}
This table is available in its entirety in a machine-readable form in the online journal. A portion is shown here for guidance regarding its form and content.
\end{table*}

%________________________________________________________________
\section*{acknowledgements}
%________________________________________________________________

We thank E. S\'anchez Blanco, M. C. C\'ardenas V\'azquez, N.
Piskunov, F. Pepe, C. Lovis, R. Cosentino, X. Dumusque, D. Staab and C. Haswell for constructive comments and discussions. We acknowledge funding from
AYA2011-30147-C03-01 by MINECO/Spain, FEDER funds/EU, and
2011-FQM-7363 of Junta de Andaluc\'ia/Spain (ZMB, PJA and CR-L); ZMB
acknowledges support from the UGR/Spain and financial funding from
FPI BES-2011-049647 of MINECO/Spain. This study is based on
observations made with the Italian Telescopio Nazionale Galileo
(TNG) operated on the island of La Palma by the Fundaci\'on Galileo
Galilei of the INAF (Istituto Nazionale di Astrofisica) at the
Spanish Observatorio del Roque de los Muchachos of the Instituto de
Astrof\'isica de Canarias. The authors thank the referee T. B\"ohm for his
suggestions that helped improved this paper.

%-----------------------------------------------------------------------------------------------------------
\bibliography{masterbibzaira}
\bibliographystyle{mn2e}

\label{lastpage}
\end{document}